 \documentclass[aps, twocolumn, showpacs, amsmath]{revtex4}
\usepackage{dcolumn}
\usepackage{bm}
\usepackage{graphicx}
\usepackage{color}
\usepackage{subfigure}
\usepackage{hyperref}
\usepackage{latexsym}
\usepackage{amsthm}
\usepackage{amssymb}

\DeclareGraphicsExtensions{.jpg,.pdf, .mps, .png, .eps, .ps, .EPS,.gif}

\begin{document}
\def\be{\begin{equation}}
\def\ee{\end{equation}}

\def\bc{\begin{center}}
\def\ec{\end{center}}
\def\bea{\begin{eqnarray}}
\def\eea{\end{eqnarray}}
\newcommand{\avg}[1]{\langle{#1}\rangle}
\newcommand{\Avg}[1]{\left\langle{#1}\right\rangle}

\def\ie{\textit{i.e.}}
\def\etal{\textit{et al.}}
\def\m{\vec{m}}
\def\G{\mathcal{G}}

\newcommand{\davide}[1]{{\bf\color{blue}#1}}
\newcommand{\gin}[1]{{\bf\color{green}#1}}

\title{Message passing theory for percolation models \\on multiplex networks with link overlap}

\author{Davide Cellai}
\affiliation{ Idiro Analytics, Clarendon House, 39 Clarendon Street, Dublin 2, Ireland\\
MACSI, Department of Mathematics and Statistics, University of Limerick, Ireland}

\author{Sergey N. Dorogovtsev}
\affiliation{Departamento de Fisica da Universidade de Aveiro, 13N, 3810-193, Aveiro, Portugal\\
A. F. Ioffe Physico-Technical Institute, 194021 St. Petersburg, Russia}

\author{ Ginestra Bianconi}
\affiliation{School of Mathematical Sciences, Queen Mary University of London, London, E1 4NS, United Kingdom}

\begin{abstract}
Multiplex networks describe a large variety of complex systems including infrastructures, transportation networks and biological systems. Most of these 
networks feature a significant link overlap. It is therefore of particular importance to characterize the mutually connected giant component in these networks. 
Here we provide a message passing theory for characterizing the percolation transition in multiplex networks with link overlap and an arbitrary number of layers $M$.
Specifically we propose and compare two message passing algorithms, that generalize the algorithm widely  used to study the percolation transition in multiplex networks without link overlap.
The first algorithm describes a directed percolation transition
and admits an epidemic spreading interpretation. The second algorithm describes  
the emergence of the mutually connected giant component,  that is the percolation transition, but does not preserve the epidemic spreading interpretation. 
We obtain the phase diagrams for the percolation and directed percolation transition in simple representative cases. 
We demonstrate that for the same multiplex network structure, in which the directed percolation transition has non-trivial tricritical points, the percolation transition 
has a discontinuous phase transition, with the exception of the trivial case in which all the 
layers completely overlap.
\end{abstract}


\maketitle
\section{Introduction}
\label{sec:introduction}

Multilayer networks \cite{PhysReports,Kivela,Goh_review} describe  complex systems formed by different interacting networks. Examples of multilayer networks are ubiquitous, ranging from infrastructures and transportation networks to cellular and brain networks \cite{Thurner,Mucha,Bullmore2009,Boccaletti,Makse,Weighted, Vito}. Characterizing the robustness of multilayer networks  is 
important for predicting the response to damage of infrastructures, transportation networks and biological networks. 
In multilayer networks, nodes from different layers are often interdependent. 
The interdependence between nodes implies that a  node is damaged if 
its interdependent nodes are damaged. 
Recently a generalized percolation process has been proposed to study the robustness of multilayer networks in 
the 
presence of interdependencies \cite{Havlin1,Dorogovtsev,Son}. 
This model 
allows to understand and to control the  
fragility of interconnected infrastructures. 
It also enables us to describe possible scenarios for   
generalized percolation processes.

Percolation on single random (locally tree-like) networks can be 
treated with 
tools 
exploiting the locally tree-like approximation \cite{crit} familiar to statistical mechanics and, alternatively,  with 
 message passing algorithms \cite{Lenka,Mezard,Weigt}.  
 Notably, the challenge of going beyond the tree-like approximations has been addressed in 
 complex 
 networks either by considering percolation on self-similar networks \cite{Marianselfsimilar} or, more recently, 
 by modifying and generalizing  message passing algorithms \cite{Radicchi_beyond}.

In single networks, the percolation transition,   
leading to the 
emergence of the giant connected component in a network, is a continuous phase transition.
A generalization of the giant connected component for multilayer networks with interdependencies between the nodes 
is called a \emph{mutually connected giant component} (MCGC) \cite{Havlin1,Dorogovtsev,Son}.
As the 
fraction of damaged nodes increases, 
a discontinuous, hybrid phase transition occurs, after which the MCGS emerges, 
and the response of the system to perturbation is characterized by large avalanches of failure events that propagate back and forth between different layers \cite{Havlin1}. 
Interestingly, the  hybrid phase transition combines a discontinuity and a critical singularity. 
The nature of this phase transition is a clear sign that multilayer networks with interdependencies  display a significant fragility with respect to random damage. 
Several other generalized percolation  
problems on 
multiplex networks have been also proposed, including competition between the layers \cite{Antagonist,Kun_q}, weak percolation \cite{Weak,baxter2016unified}, generalized $k$-core percolation \cite{Kcore}, 
percolation on directed multiplex networks \cite{Doro_directed}, spanning connectivity \cite{Guha2016}, and bond percolation \cite{Bond}. 

The emergence of the MCGC has been studied on a variety of multilayer structures  including multiplex networks \cite{Havlin1,Dorogovtsev,Son,Goh,HavlinEPL,Havlin2,Stanleyint,Cellai2016} and networks of networks \cite{Gao1,BD1,BD2}.
Networks of networks are multilayer networks formed by different networks (layers), where the nodes of different  networks might be related by interdependencies. The percolation transition in these networks is significantly affected by the way the links implying interdependencies are placed, and, moreover,  
instead of single, there may be  
multiple 
transitions \cite{BD1,BD2}.

Multiplex networks describe a large variety of complex systems and constitute a well controlled setting to study the interplay between structure and dynamics in multilayer networks. They  are formed by a set of $N$ nodes interacting via $M$ different layers. Each node has a replica node in each layer, and each layer is a distinct network 
for the replica nodes in that layer. The interdependencies in multiplex networks are usually placed between the 
replica nodes from different layers. 
The percolation phase transition describing the emergence of the MCGC in multilayer networks, with layers formed by random networks with given degree distributions, has been fully characterized as a discontinuous and 
hybrid transition \cite{Havlin1,Dorogovtsev}. The  phase transition  remains hybrid and discontinuous  in 
the 
presence of correlations in the degrees of replica nodes \cite{Goh} but can become a continuous  in the case of partial interdependence \cite{HavlinEPL,Havlin2,Stanleyint} or if some nodes are not active (not connected) in each layer \cite{Cellai2016,Vito}. 
Interestingly,   
these results  can be obtained using a locally tree-like approximation, or equivalently,  a message passing algorithm which admits  an epidemic spreading interpretation \cite{Son,Kabashima}.

Numerous multilayer networks have a  significant link overlap \cite{PRE,Thurner,Boccaletti,Weighted},  
which explains the need to 
explore the percolation transition 
on this type of correlated multilayer structures \cite{note0}. 
 Recently, two approaches were used to describe 
the transition in duplex networks ({\ie} networks formed by $M=2$ layers) with link overlap. 
The first approach consists of a coarse-grained description of the multiplex network in terms of supernodes \cite{supernodes,Goh_comment}. The  second  approach is instead based only on a traditional local tree-like approximation \cite{Baxter2016}. 
Interestingly, it turns out that a message passing algorithm that admits  an epidemic spreading interpretation \cite{Cellai2013,Radicchi}, inspired by the algorithm originally proposed for multiplex network without link overlap,  does not capture the MCGC \cite{supernodes, Goh_comment,Baxter2016}, but instead characterizes  a new type of  directed percolation. 
This process can be  interpreted  as a variation of a bootstrap percolation dynamics \cite{bootstrap,Weak} or a as  the viability percolation problem \cite{Goh_comment} in the limit in which the resource nodes are vanishing.
Here we call this dynamical process  directed percolation and its  order parameter  \emph{directed mutually connected giant component} (DMCGC) to distinguish it from the MCGC.
The choice of this terminology is due to the fact that  we want to highlight the directed nature of the underlying process, and the connection to epidemic spreading \cite{Azimi} processes nevertheless we want to clarify that the links of the underlying multiplex network do not have an intrinsic directionality.

Our unified approach to percolation and directed percolation is directly applicable to multiplex network with link overlap and arbitrary number of layers $M$. This approach  is used to fully characterize and  compare the percolation transitions and the directed percolation transitions on ensembles of random multiplex networks.
We show that while the directed percolation transition in multiplex networks with link overlap can have non-trivial tricritical points, the percolation transition on 
this type of multiplex networks is always hybrid and discontinuous with the sole trivial exception 
in that all the layers of a multiplex network completely overlap with each other. 

Message passing algorithms are attracting increasing 
attention in network theory. 
They were used 
to characterize the structure of single 
networks \cite{Radicchi} or epidemic spreading in temporal multi-slices networks \cite{Colizza}, and to detect the driver nodes 
controlling a network \cite{Control}.
This work shows that a new class of  message passing algorithms can be used to investigate the structure of multiplex networks with link overlap, allowing 
to characterize both the MCGC and DMCGC in locally tree-like networks. 

Interestingly, the DMCGC  is related to directed cooperative epidemic spreading in multiplex networks, while the MCGC characterizes the response of the interdependent multiplex network structure to external damage. 

The DMCGC model could provide therefore an ideal setting to extend models of cooperative contagion studied on single networks \cite{Fakhteh} to multilayer networks.
 
In Sec.~\ref{sec:no-overlap} and \ref{sec:giant-component} we recall the configuration model for multiplex networks and the definition of mutually connected giant component (MCGC).
In Sec.~\ref{sec:percolation-no-overlap} we report the message passing theory for calculating the MCGC in multiplex networks without link overlap,  using the formalism  
of Ref.~\cite{PRE}.
In Sec.~\ref{sec:with-overlap} we present the multilink definitions in the case of link overlap, using the formalism 
of Ref.~\cite{Cellai2013}.
In Sec.~\ref{sec:two-mutual-components}, we clarify the extension of the message passing approach to multiplex networks with link overlap. We distinguish between two possible extensions, leading to a directed mutually connected giant component (DMCGC) and a mutually connected giant component (MCGC), respectively.
In Sec.~\ref{sec:DMCGC} we briefly describe the results obtained with DMCGC.
The main novel contribution of this paper is in Sec.~\ref{sec:MCGC}, where we define the message passing approach for MCGC in the presence of link overlap, and present results for a few particular cases.
Finally, Sec.~\ref{sec:conclusions} reports our conclusions.

\begin{figure*}[htb]
\begin{center}
	\includegraphics[width=2.0\columnwidth]{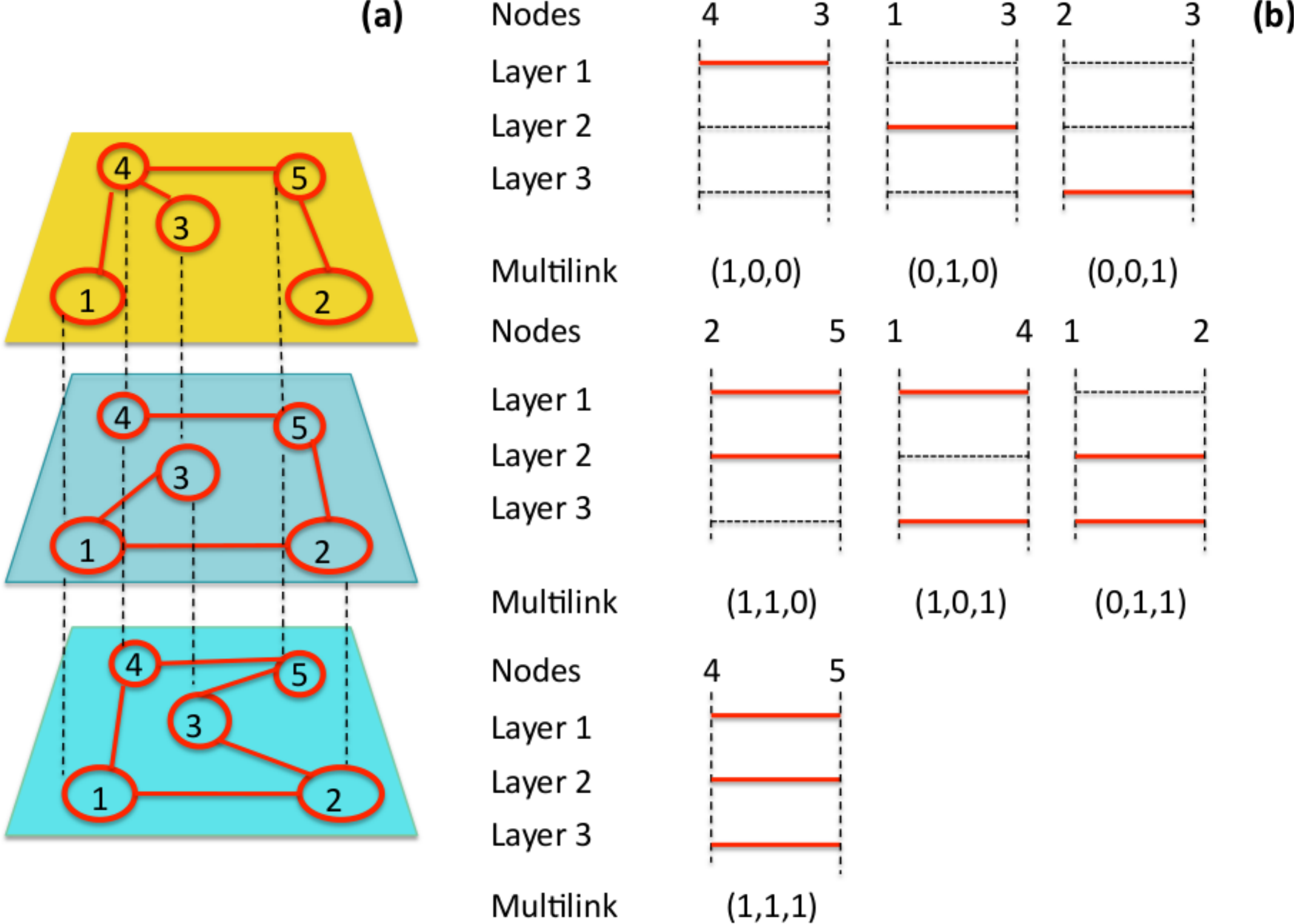}
	\caption{(Color online) A multiplex network with $M=3$ layers and link overlap is shown in panel (a). 
	In panel (b) the different types of non-trivial multilinks connecting the nodes are listed.}
	\label{Figure1}
\end{center}
\end{figure*}

\section{Multiplex networks without link overlap}
\label{sec:no-overlap}

A multiplex network $\vec{\G}=(G_1,G_2,\ldots G_M)$ is formed by a set of $N$ nodes $i=1,2,\ldots,N$ interacting through $M$ layers with each layer $\alpha=1,2,\ldots, M$  formed by a  distinct network $G_{\alpha}$. Every node $i$ has $M$ replica nodes, one for each layer $\alpha$ indicating the node identity in layer $\alpha$. Replica nodes are connected pairwise by {\it interlinks}. Figure~\ref{Figure1} shows an example of a multiplex network with $M=3$ layers. 
A network $G_{\alpha}$ in layer $\alpha$ connects the $N$ replica nodes in this layer. This network is fully described by the adjacency matrix ${\bf a}^{[\alpha]}$. The matrix element $a_{ij}^{[\alpha]}=0,1$ of the adjacency matrix ${\bf a}^{[\alpha]}$ indicates whether  node $i$ is connected to node $j$ in layer $\alpha$ ($a_{ij}^{[\alpha]}=1$) or  not  ($a_{ij}^{[\alpha]}=0$).

In a multiplex network, we 
define the total overlap  ${O}^{[\alpha,\alpha']}$ \cite{PRE}  of the links in layer $\alpha$ and layer $\alpha'$ as the total number of pairs of nodes connected both in layer $\alpha$ and layer $\alpha'$, {\ie}
\bea
{O}^{[\alpha,\alpha']}=\sum_{i,j}a^{[\alpha]}_{ij}a^{[\alpha']}_{ij}.
\eea
Additionally we 
define \cite{PRE} for each node $i$ the local overlap  of the links in layer $\alpha$ and layer $\alpha'$ as the total number of neighbors of node $i$ which are simultaneously neighbors in layer $\alpha$ and in layer $\alpha'$, {\ie}
\bea
o_i^{[\alpha,\alpha']}=\sum_{j=1}^N a^{[\alpha]}_{ij}a^{[\alpha']}_{ij}.
\label{oi}
\eea
We stress that most real multiplex networks have a significant total and local overlap of their links \cite{Thurner, Boccaletti, Weighted}.

 The first natural approach to construct an ensemble of random multiplex networks, is to generate each layer independently.
For this, we draw  the sequence of the degrees $\{k_i^{[1]},k_i^{[2]}, \ldots, k_i^{[M]}\}$  of each node $i$  from 
a given degree distribution $P(\{k^{[\alpha]}\})$.
If the degrees of each individual node in different layers are uncorrelated, the degree distribution $P(\{k^{[\alpha]}\})$ factorizes as 
\bea
P(\{k^{[\alpha]}\})=\prod_{\alpha}P^{[\alpha]}(k^{[\alpha]}),
\label{f}
\eea
where $P^{[\alpha]}(k^{[\alpha]})$ is the degree distribution in layer $\alpha$.
Finally, having assigned to each node $i$ the degree $k_i^{\alpha}$ in every layer $\alpha$,   we can construct  a multiplex network in which each layer is a random graph with given degree sequence $\{k_1^{[\alpha]},k_2^{[\alpha]}\ldots, k_N^{[\alpha]}\}$, {\ie} we consider a random  multiplex ${\vec{\G}}$ chosen with probability 
\bea
P({\vec {\G}})=\prod_{\alpha=1}^M \prod_{i=1}^N \delta(k_i^{[\alpha]},\sum_j a_{ij}^{[\alpha]}),
\label{ens}
\eea
where $\delta(x,y)$ here and in the following indicates the Kronecker delta.
As long as the layers are formed by sparse networks, and the number of layers is much smaller than the number of nodes, {\ie} $M\ll N$,  the multiplex network constructed in this way has a negligible total link overlap $O^{[\alpha,\alpha']}$ between any two layers $\alpha$ and $\alpha'$, and a negligible local overlap $o_i^{[\alpha,\alpha']}$ also \cite{PRE}.  Since the link overlap is a general property of multiplex networks, 
one should 
consider multiplex network models able to reproduce this structural feature. Such models have been introduced in \cite{PRE} and will be discussed in Sec.~\ref{sec:with-overlap}.

For the sake of simplicity we will assume throughout the paper that each node is active ({\ie} connected) in every layer. This assumption can be relaxed. We refer the interested reader to Refs.~\cite{Vito, Cellai2016, Guha2016} where this case and its implications for the percolation transition, including asymptotic behavior in case of a large number of layers, are discussed in detail.

\section{Percolation on multiplex networks and mutually connected giant component}
\label{sec:giant-component}

In this paper  we 
consider the robustness of  multiplex networks in the presence of interdependencies.

Following Ref.~\cite{Havlin1} we 
assume that each interlink indicates an interdependency between the linked replica nodes. This interdependencies  imply that if a replica node is damaged, then all the interdependent replica nodes in the other layers are damaged. The robustness of the multiplex network is monitored by 
the response to an external initial damage performed on a set of nodes of the network.
The variables $\{s_i \}$, where ${i=1,2,\ldots, N}$, fully characterize the inital damage to the network, as each variable $s_i$   indicates whether  node $i$  has been  initially damaged ($s_i=0$) or not ($s_i=1$).
The  multiplex network responds non-linearly to this damage as it can be 
quantifying by the size of its mutually connected giant component (MCGC).
The MCGC  has been defined in \cite{Havlin1}  as a  generalization of the giant component of single networks.  
This is the component that remains after the damage propagates back and forth between the layers. The original algorithm that defines the MCGC is the following: 

\begin{itemize}

\item[(i)]   
the giant component of each layer $\alpha$ is determined, evaluating the effect of the damaged nodes in each single layer;

\item[(ii)] 
each node that has at least a replica node not in the giant component of its proper layer is damaged, {\ie}  all its replica nodes are damaged   due to the interdependencies existing between them;

\item[(iii)]  
If there are no new damaged nodes the algorithm stops, otherwise it proceed, starting again from step~(i).

\end{itemize}
At the end of the iteration the nodes that are not damaged by the iterative process form the MCGC. The size $NS$ of the MCGC is given by the number of nodes remaining undamaged by this process.
If  the initial damage $\{s_i\}$ has probability distribution 
\bea
\pi(\{s_i\})=\prod_{i=1}^N \left[p s_i+(1-p)(1-s_i)\right],
\label{Ps}
\eea 
{\ie} initially each node is damaged independently with probability $1-p$,
we observe a phase transition with the order parameter $S$ and the control parameter $p$.

\section{Percolation in multiplex networks without link overlap}
\label{sec:percolation-no-overlap}

\subsection{The message passing algorithm}

On a locally tree-like multiplex network without link overlap, the MCGC can be found by using a suitable message passing algorithm. This algorithm has been first proposed by Son et al. \cite{Son}.  According to this algorithm, nodes send messages along their  links to neighbor  nodes. Each message sent by a node $i$ to a node $j$   indicates whether node $i$  belongs to the MCGC also in absence of the link $(i,j)$.
In particular the message $\sigma_{i\to j}^{\alpha}$ that node $i$ send to a neighbor node $j$ in layer $\alpha$ is equal to one ($\sigma_{i\to j}^{\alpha}=1$) if the following conditions are met:

\begin{itemize}

	\item[(a)] node $i$ is not initially damaged, {\ie} $s_i=1$;
	
	\item[(b)] node $i$ belongs to the  mutually connected giant component even if the link  between node $j$ and node $i$ is removed from the multiplex, {\ie} for every layer $\alpha'=1,2\ldots, M$ node $i$ receives at least one positive message $\sigma_{\ell\to i}=1$ from  nodes  $\ell \neq j$ that are neighbors of node $i$  in layer $\alpha'$.
	
\end{itemize}

If these conditions are not met, then  $\sigma_{i\to j}^{\alpha}=0$. 
These messages determine whether a node $i$ belongs ($\sigma_i=1$) or not ($\sigma_i=0$) to the MCGC. In fact node $i$ belongs to the MCGC ($\sigma_i=1$) if and only if  

\begin{itemize}

	\item[(a)] node $i$ is not initially damaged;
	
	\item[(b)]  node $i$ receives at least one positive message $\sigma_{\ell\to i}=1$ from a  neighbor $\ell$ of node $i$ in every  layer $\alpha$. 
	
\end{itemize}

These two algorithms directly translate in the message passing equations
\bea
\sigma_{i\to j}^{\alpha}&=&s_i\left[1-\prod_{\ell \in N_{\alpha}(i)\setminus j}(1-\sigma_{\ell\to i}^{\alpha})\right]\nonumber \\
&&\times \prod_{\alpha'\neq \alpha}\left[1-\prod_{\ell \in N_{\alpha'}(i)}(1-\sigma_{\ell\to i}^{\alpha'})\right].\nonumber \\
\sigma_{i}&=&s_i \prod_{ \alpha=1}^M\left[1-\prod_{\ell \in N_{\alpha}(i)}(1-\sigma_{\ell\to i}^{\alpha})\right].
\eea
where $N_{\alpha}(i)$ indicates the set of neighbors of node $i$ in layer $\alpha$.
Let us consider a random multiplex network taken with probability given by Eq.~(\ref{ens}) and a random realization of the initial damage described by the probability given by Eq.~(\ref{Ps}). The average message in layer $\alpha$, $S'_{\alpha}=\Avg{\sigma_{i\to j}}$ and the (relative) average number of nodes in the MCGC,  $S=\Avg{\sigma_i}$ are given by 
\bea
S&=&p\sum_{\{k^{\alpha}\}}P(\{k^{\alpha}\})\prod_{\alpha=1}^M[1-(1-S'_{\alpha})^{k_{\alpha}}],
\nonumber 
\\
S'_{\alpha}&=&p\sum_{\{k^{\beta}\}}\frac{k^{\alpha}}{\Avg{k^{\alpha}}}P(\{k^{\beta}\})[1-(1-S'_{\alpha})^{k^{\alpha}-1}]
\nonumber 
\\
&&\times\prod_{\alpha'\neq \alpha}[1-(1-S'_{\alpha'})^{k^{\alpha'}}].
\eea
If there are no correlations between the degrees of a node in different layers, 
and so the degree distribution $P(\{k^{[\alpha]}\})$ follows Eq.~(\ref{f}), then we have 
\bea
S&=&p\prod_{\alpha=1}^M[1-G_0^{[\alpha]}(1-S'_{\alpha})],
\nonumber 
\\
S'_{\alpha}&=&p[1-G_1^{[\alpha]}(1-S'_{\alpha})]\prod_{\alpha'\neq \alpha}[1-G_0^{[\alpha']}(1-S'_{\alpha'})].
\label{NL}
\eea
Here the  generating functions $G_0^{[\alpha]}(z)$
and $G_1^{[\alpha]}(z)$ of the degree distribution $P^{[\alpha]}(k)$ of layer $\alpha$  are given by 
\bea
G_0^{[\alpha]}(z)&=&\sum_{k}P^{[\alpha]}(k)z^k, \nonumber \\
G_1^{[\alpha]}(z)&=&\sum_{k}\frac{k}{\avg{k^{[\alpha]}}}P^{[\alpha]}(k)z^{k-1}.\label{SLMCGC}
\eea

\subsection{The case of equally distributed Poisson layers}

In the case of equally distributed Poisson layers with average degree $c$, we have 
\bea
P^{[\alpha]}(k)=\frac{1}{k!}c^ke^{-c}
\eea
for every layer $\alpha=1,2,\ldots, M$.
Then, using Eqs.~(\ref{NL}), 
one can show that $S'_{\alpha}=S$ for every layer $\alpha$, and $S$ is determined by the equation 
\bea
S=p\left(1-e^{-cS}\right)^M.
\eea
By setting $S/p=x$, this equation reduces to $h_{cp}(x)=0$, where the function $h_{cp}(x)$ is 
\bea
h_{cp}(x)=x-(1-e^{-cpx})^M=0.
\eea
This equation has always the trivial solution $x=0$. In addition, a non-trivial solution $x>0$ indicating the presence of the MCGC, emerges at a hybrid discontinuous transition at $x=x_c$, $cp=cp_c$ determined by the equations 
\bea
h_{cp}(x_c)&=&0,\nonumber \\
\left.\frac{dh_{cp}(x)}{dx}\right|_{x=x_c}&=&0.
\eea 
For $M=2$ this yields the discontinuous hybrid transition for $cp_c\simeq 2.4554$, $x_c=S_c/p \simeq 0.5117$ \cite{Havlin1,Son,Dorogovtsev}. 
For $M=3$ this yields the discontinuous hybrid transition for $cp_c\simeq 3.0891$, $x_c=S_c/p \simeq 0.6163$.

\section{Multiplex networks with link overlap}
\label{sec:with-overlap}

The vast majority of multiplex networks in  infrastructures, transport, social  and collaboration networks  are characterized by significant link overlap \cite{Thurner, Boccaletti, Weighted}.  Therefore it is of crucial importance to determine the robustness of multiplex networks in presence of this structural feature.
In order to model multiplex networks with link overlap, the notion of {\it multilinks} \cite{PRE,Weighted, Cellai2013} turns out to be extremely  useful.
Two nodes $i$ and $j$ are connected by a {\it multilink} $\vec{m}=(m_1,m_2,\ldots, m_M)$ with $m_{\alpha}=0,1$, if and only if they are linked in every layer $\alpha$ for which $m_{\alpha}=1$ and they are not linked in every layer $\alpha$ for which $m_{\alpha}=0$ (see Fig.~\ref{Figure1} for a graphical  description of multilinks).
We distinguish between the non-trivial multilinks $\vec{m}\neq \vec{0}$ and the trivial multilink $\vec{m}=\vec{0}$ indicating the absence of any sort of link between the two nodes.

Using the concept  of multilinks  one can define  multi-adjacency matrices ${\bf A}^{\vec{m}}$ whose element $A_{ij}^{\vec{m}}$ indicates whether  node $i$ is connected to node $j$ by a multilink $\vec{m}$ ($A_{ij}^{\vec{m}}=1$) or not $(A_{ij}^{\vec{m}}=0)$.
The multi-adjacency matrices encode the same information encoded in the adjacency matrices ${\bf a}^{\alpha}$ and the matrix elements $A_{ij}^{\vec{m}}$ can consequently  be expressed as a function of the matrix elements $a_{ij}^{[\alpha]}$ as 
 \bea
 A_{ij}^{\vec{m}}=\prod_{\alpha=1}^M[m_{\alpha}a_{ij}^{[\alpha]}+(1-m_{\alpha})(1-a_{ij}^{[\alpha]})].
 \eea
 The multi-adjacency matrices are not independent in 
 in the sense that they satisfy
 \bea
 \sum_{\vec{m}}A_{ij}^{\vec{m}}=1
 \eea
 for every pair of nodes $(i,j)$ of the multiplex network.  Having introduced the multi-adjacency matrices it is straightforward to define the multidegrees \cite{PRE,Weighted, Cellai2013}. 
 The  {\it multidegree} $\vec{m}$ of node $i$ indicated as $k_i^{\vec{m}}$ is the sum of rows (or column) of the multi adjacency matrix ${\bf A}^{\vec{m}}$, {\ie} 
 \bea
 k_i^{\vec{m}}=\sum_j A_{ij}^{\vec{m}}.
 \eea
 Therefore, the multidegree $k_i^{\vec{m}}$ indicates the number of nodes linked to node $i$ by a multilink $\vec{m}$.
 As an example, consider a multiplex network (duplex) formed by two layers. 
 Using the adjacency matrices of elements $a_{ij}^{[\alpha]}$, the multidegrees of node $i$ are given by 
 \bea
 k_i^{(1,1)}&=&\sum_ja_{ij}^{[1]}a_{ij}^{[2]},\nonumber \\
 k_i^{(1,0)}&=&\sum_ja_{ij}^{[1]}(1-a_{ij}^{[2]}),\nonumber \\
 k_i^{(0,1)}&=&\sum_j(1-a_{ij}^{[1]})a_{ij}^{[2]},\nonumber \\
 k_i^{(0,0)}&=&\sum_j(1-a_{ij}^{[1]})(1-a_{ij}^{[2]}).
 \eea
 From the explicit expression of the multidegree $k_i^{(1,1)}$ it is evident that this quantity is given by the local overlap $o_i^{[1,2]}$ defined in Eq.~(\ref{oi}), {\ie} 
 \bea
 k_i^{[1,1]}=o_i^{[1,2]}.
 \eea
 Therefore $k_i^{[1,1]}$ indicates the number of neighbors of node $i$ that are simultaneously neighbor in layer $1$ and layer $2$. On the contrary, the multidegree $k_i^{[1,0]},$/$k_i^{[0,1]}$ indicate respectively the number of neighbors of node $i$ that are neighbor in layer 1/(layer 2) but not in layer 2/(layer 1). Finally the multidegree $k_i^{[0,0]}$ indicates the total number of nodes that are not connected to node $i$ in any layer.
 
 In general, for arbitrary (but finite) number of layer $M$, the multidegrees of a node give a complete, local information about the link overlap in different layers.
 
 Naturally, since the multiadjacency matrices are not independent, also the multidegrees of a node are not all independent, and we have
 \bea
 \sum_{\vec{m}}k_i^{\vec{m}}=N,
 \eea
 for every node $i$, 
 which can also be written as 
 \bea
 k_i^{\vec{0}}=N-\sum_{\vec{m}\neq \vec{0}}k_i^{\vec{m}}.
 \eea
 In a sparse multiplex network the non-trivial multidegrees $k^{\vec{m}}$ with  $\vec{m}\neq \vec{0}$ have finite average $\Avg{k^{\vec{m}}}$. 
 
 Random multiplex networks with a given distribution of the multidegree sequence $P(\{k^{\vec{m}}\})$ provide the 
 easiest way to generate multiplex networks with a controlled link overlap. In order to do this we first draw the sequence $\{k_i^{\vec{m}}\}$ of multidegrees of each node $i$ from the multidegree distribution $P(\{k^{\vec{m}}\})$. To each node $i$ we associate $k_i^{\vec{m}}$ stubs of type $\vec{m}$ and finally we match pairwise stubs of the same multilink type. In this way the probability that node $i$ and node $j$ are connected by a multilink $\vec{m}\neq \vec{0}$ is  given by \cite{PRE}
 \bea
 p_{ij}^{\vec{m}}=\frac{k_i^{\vec{m}}k_j^{\vec{m}}}{\avg{k^{\vec{m}}}N},
 \eea 
 as long as the multidegrees have the natural structural cutoff, {\ie} 
 \bea
 k_i^{\vec{m}}<\sqrt{\Avg{k_i^{\vec{m}}}N},\eea  for every multilink $ \vec{m}\neq \vec{0}$.
In this ensemble the probability $P(\vec{\G})$ of a multiplex $\vec{\G}$ is given by
\bea
P(\vec{\G})=\prod_{ij}\prod_{\vec{m}\neq \vec{0}}\delta\left({k_i^{\vec{m}},\sum_jA_{ij}^{\vec{m}}}\right).
\label{PGM}
\eea
Eventually, the multidegrees of a given node can  be uncorrelated, {\ie} the multidegree distribution  factorizes 
\bea
{P}(\{k^{\vec{m}}\})=\prod_{\vec{m}\neq \vec{0}}P^{\vec{m}}(k^{\vec{m}}),
\label{UPGM}
\eea
where $P^{\vec{m}}(k)$ is the distribution of multidegrees $k^{\vec{m}}=k$ with $\vec{m}\neq \vec{0}$.

\section{Mutually connected component and Directed mutually connected component in multiplex networks with link overlap}
\label{sec:two-mutual-components}

\begin{figure*}[htb]
\begin{center}
	\includegraphics[width=2.0\columnwidth]{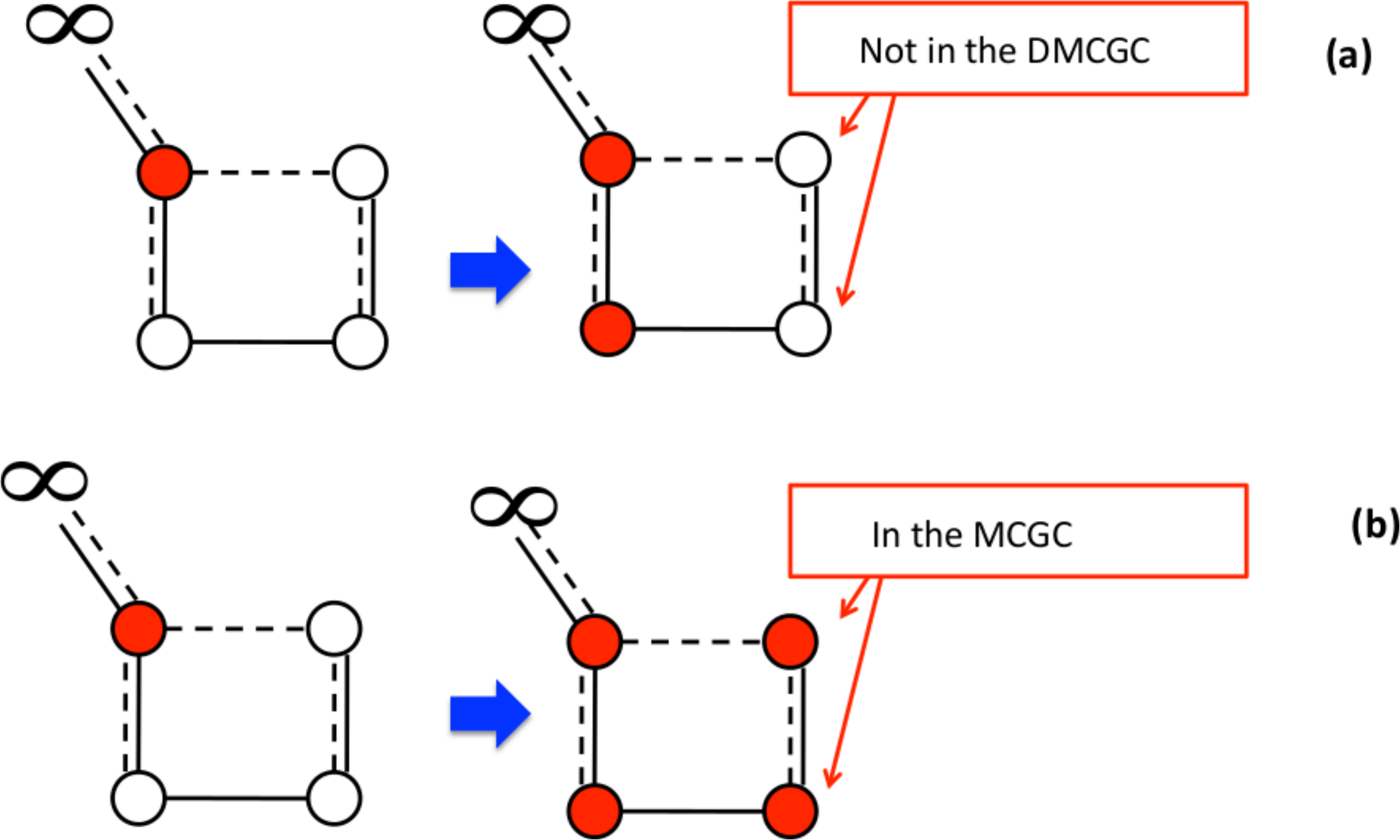}
	\caption{(Color online) A multiplex network with link overlap 
demonstrating that the DMCGC is not 
equivalent to the MCGC. Here the multiplex network has $M=2$ layers corresponding to the networks formed by links indicated respectively with solid and dashed lines. In panel (a) we assume that one node is connected to the DMCGC. By applying the message passing algorithm described in Sec.~\ref{sec:DMCGC}, we observe that two nodes of the network do not belong to the DMCGC. In panel (b) we consider the same multiplex network configuration but this time we assume that a single node is connected to the MCGC. By applying the message passing algorithm described in Sec.~\ref{sec:MCGC} we observe that all the nodes of this network belong to the MCGC.}
	\label{Figure2}
\end{center}
\end{figure*}

The message passing algorithm discussed in Sec.~\ref{sec:percolation-no-overlap}  can be  extended and 
used to explore the structure of  locally tree-like  multiplex networks with link overlap in different ways. 
We consider two extensions. 
With the first algorithm (Sec.~\ref{sec:DMCGC}), one characterizes the directed mutually connected giant component (DMCGC), with the second algorithm (Sec.~\ref{sec:MCGC}) one characterizes the mutually connected giant component (MCGC).
Both algorithms reduce to the algorithm studied in Sec.~\ref{sec:percolation-no-overlap} in the absence of link overlap. 
Moreover, both algorithms reduce to percolation on a single network  in the presence of complete overlap of all the layers.
The algorithm \cite{Cellai2013} that calculates the DMCGC has an  epidemic spreading interpretation and an inherent directed character.
In this epidemic spreading interpretation, we assume that a different disease propagate in each layer of the multiplex networks and that a node is infected ({\ie} it sends a positive message to a downstream node) only if it is in contact to at least an infected upstream neighbor in every layer $\alpha=1,2,\ldots, M$. The set of nodes that become infected are the nodes in the DMCGC.
It  has been shown that this algorithm determines a \emph{proper} subset of the nodes that are in the MCGC as soon as there is link overlap \cite{Goh_comment}. 
For example, 
for the network in Fig.~\ref{Figure2}, 
all the nodes of the drawn networks belong to the MCGC, but, according to the 
message passing algorithm with the epidemic spreading interpretation, two nodes remain uninfected. That is, while these two nodes belong to the MCGC, they do not belong to the DMCGC \cite{Goh_comment}.

It has been 
debated if 
a message passing algorithm allows to describe 
the mutually connected giant component in multiplex networks 
with an arbitrary number of layers $M$. Recently a traditional tree-like approximation 
was successfully used to characterize the mutually connected giant component in a multiplex networks with link overlap and $M=2$ \cite{Baxter2016}.
Here we show that it is possible to extend these results to multiplex network with arbitrary number of layers $M$ and link overlap by using  a message passing algorithm combined with the use of multilinks.

In the following we treat and compare two different types of message passing algorithms: one for directed percolation and the other for percolation, which can detect the nodes belonging respectively to the DMCGC and to the MCGC.
Applying these algorithms one can study the critical properties of the two percolation transitions and observe significant changes in the phase diagrams of these problems. 
In Sec.~\ref{sec:DMCGC}, we describe directed percolation and then, in Sec.~\ref{sec:MCGC}, we  calculate the size of the MCGC with the message passing approach, in multiplex networks with link overlap. 

\section{Directed percolation of multiplex networks}
\label{sec:DMCGC}

 \subsection{The message passing algorithm}
 
In a locally  tree-like multiplex with link overlap a simple extension of the message passing  
from Sec.~\ref{sec:percolation-no-overlap} determines the set of nodes belonging to the directed mutually connected giant component (DMCGC) \cite{Cellai2013}.
Let $s_i=0,1$ indicate if a node $i$ is removed or not from the network and let   $\sigma_i=0,1$ be the indicator function that the node $i$ is in the  DMCGC. The value of $\sigma_i$ is determined by the ``messages" that the neighboring nodes send to node $i$. We denote the message sent from node $i$ to node $j$ as $\sigma_{i\to j}^{\vec{m}^{ij}}$. The value of this   is set to one $\sigma_{i\to j}^{\vec{m}^{ij}}=1$ if and only if the following three conditions are satisfied:
\begin{itemize}
	\item[(a)] node $j$ is a neighbor of node $i$ with a multilink $\vec{m}^{ij}$ connecting them such that $\sum_{\alpha}m_{\alpha}^{ij}>0$;
	\item[(b)] node $i$ is not initially damaged, {\ie} $s_i=1$;
	\item[(c)] node $i$ belongs to the directed mutually connected giant component even if the multilink $\vec{m}^{ij}$ between node $i$ and node $j$ is removed from the multiplex, {\ie}  node $i$ receives at least one positive message $\sigma_{\ell\to i}^{\vec{m}^{\ell i}}=1$ from a nearest neighbor $\ell \neq j$ in  every layer $\alpha$.
\end{itemize}
If any of these conditions is not satisfied then the messages is zero, {\ie} $\sigma_{i\to j}^{\vec{m}^{ij}}=0$.

Additionally, node $i$ is in the DMCGC ($\sigma_i=1$) if the following conditions are satisfied: 
\begin{itemize}
	\item[(a)]  node $i$ is not initially damaged
	\item[(b)] for every layer $\alpha$ node $i$ receives at least one positive message $\sigma_{\ell\to i}^{\vec{m}^{\ell i}}=1$ from a  neighbors $\ell$ in  layer $\alpha$. 
\end{itemize}
This algorithm directly translates into the following message passing equations for   $\sigma_i$ and $\sigma_{i\to j}^{\vec{m}^{ij}}$: 
\bea
	\sigma_i = s_i \prod_{\alpha=1}^M\left[1-\prod_{j\in N_{\alpha}(i)}(1-\sigma_{j\to i}^{\vec{m}^{ij}})\right],
	\label{sigma-1}\\\
	\sigma_{i\to j}^{\vec{m}^{ij}} = s_i \prod_{\alpha=1}^M\left[1-\prod_{\ell \in N_{\alpha}(i) \setminus j}(1-\sigma_{\ell \to i}^{\vec{m}^{\ell i}})\right],
	\label{sigma-ijm-1}
\eea
where $N_{\alpha}(i)$ indicates the set of neighboring nodes of node $i$ in layer $\alpha$.
Let us consider a random realization of the initial damage drawn from the probability distribution given by Eq.~(\ref{Ps}) and a random realization of the multiplex network with link overlap chosen with probability given by Eq.~(\ref{PGM}). The average message $S_{\vec{n}}=\Avg{\sigma_{i\to j}^{\vec{m}^{ij}}}$ along a generic multilink $\vec{m}^{ij}=\vec{n}$ and the average number of nodes in the DMCGC $S=\avg{\sigma_i}$ are respectively given by  (see Ref.~\cite{Cellai2013} for the details of the derivation)
\bea
	S_{\vec{n}}&=&p\sum_{\{k^{\vec{m}}\}}\frac{k^{\vec{n}}}{\Avg{k^{\vec{n}}}}P(\{k^{\vec{m}}\})\sum_{\vec{r}}(-1)^{\sum_{\alpha=1}^M r_{\alpha}} 
	\nonumber \\
	&&\times \left[(1-S_{\vec{n}})^{k^{\vec{n}}-1}\right]^{f({\vec{n},\vec{r})}}\ \prod_{
		\begin{subarray}{c}
        			\vec{m}|\\
			\sum_{\alpha}m_{\alpha}r_{\alpha}>0\\
			\vec{m}\neq \vec{m}
      		\end{subarray}
	}(1-S_{\vec{m}})^{k^{\vec{m}}},
	\nonumber \\
	S &=& p\sum_{\{k^{\vec{m}}\}}P(\{k^{\vec{m}}\})\sum_{\vec{r}}(-1)^{\sum_{\alpha=1}^M r_{\alpha}} 
	\nonumber \\
	&&\times\prod_{
		\begin{subarray}{c}
        			\vec{m}\\
			\sum_{\alpha}m_{\alpha}r_{\alpha}>0
      		\end{subarray}
	}(1-S_{\vec{m}})^{k^{\vec{m}}},
\eea
where   $\vec{r}=(r_1,r_2,\ldots, r_{\alpha},\ldots r_M)$ with $r_{\alpha}=0,1$ and $f(\vec{n},\vec{r})=1$ if $\sum_{\alpha}r_{\alpha}n_{\alpha}>0$ and   $f(\vec{n},\vec{r})=0$ otherwise (see Ref.~\cite{Cellai2013} for the details of the derivation).
For uncorrelated multidegrees of the nodes, when the distribution $P(\{k^{\vec{m}}\})$ follows Eq.~(\ref{UPGM}), these equations read
\bea
	S_{\vec{n}}&=&p\sum_{\{k^{\vec{m}}\}}\sum_{\vec{r}}(-1)^{\sum_{\alpha=1}^M r_{\alpha}}\left[G^1_{\vec{n}}(1-S_{\vec{n}})\right]^{f({\vec{n},\vec{r})}} 
	\nonumber \\
	&&\times \prod_{
		\begin{subarray}{c}
        			\vec{m}\\
			\sum_{\alpha}m_{\alpha}r_{\alpha}>0\\
			\vec{m}\neq \vec{n}
      		\end{subarray}
	}G^0_{\vec{m}}(1-S_{\vec{m}}),\label{DMCG}\\
	S &=& p\sum_{\{k^{\vec{m}}\}}\sum_{\vec{r}}(-1)^{\sum_{\alpha=1}^M r_{\alpha}}\prod_{
		\begin{subarray}{c}
        			\vec{m}\\
			\sum_{\alpha}m_{\alpha}r_{\alpha}>0
      		\end{subarray}
	}G^0_{\vec{m}}(1-S_{\vec{m}}). 
	\nonumber
\end{eqnarray}
where $\vec{r}$ and $f(\vec{n},\vec{r})$ have  the same definition as above, and  the generating function $G_0^{\vec{m}}(z)$ and $G_1^{\vec{m}}(z)$ are given by 
\bea
G_0^{\vec{m}}(z)&=&\sum_{k}P^{\vec{m}}(k)z^k, \nonumber \\
G_1^{\vec{m}}(z)&=&\sum_{k}\frac{k}{\Avg{k^{\vec{m}}}}P^{\vec{m}}(k)z^{k-1}.
\eea
Note that this algorithm and therefore Eqs.~(\ref{DMCG}) reduce to the Eqs.~(\ref{SLMCGC}) found in Sec.~\ref{sec:percolation-no-overlap} in the absence of link overlap, {\ie} where the only non-trivial multilinks are the ones with $\sum_{\alpha}m_{\alpha}=1$.

Here we show two simple examples of  
how this scheme works 
in practice. 
These examples have been already discussed in~\cite{Cellai2013} and checked against simulation results in \cite{Goh_comment} but we report them here to demonstrate the difference between the equation determining the DMCGC and the one for the MCGC that we calculate for the same multiplex ensembles in Sec.~\ref{m2} and Sec.~\ref{m3}.

\subsection{ Two Poisson layers with Overlap}

\begin{figure}[htb]
\begin{center}
	\includegraphics[width=1.0\columnwidth]{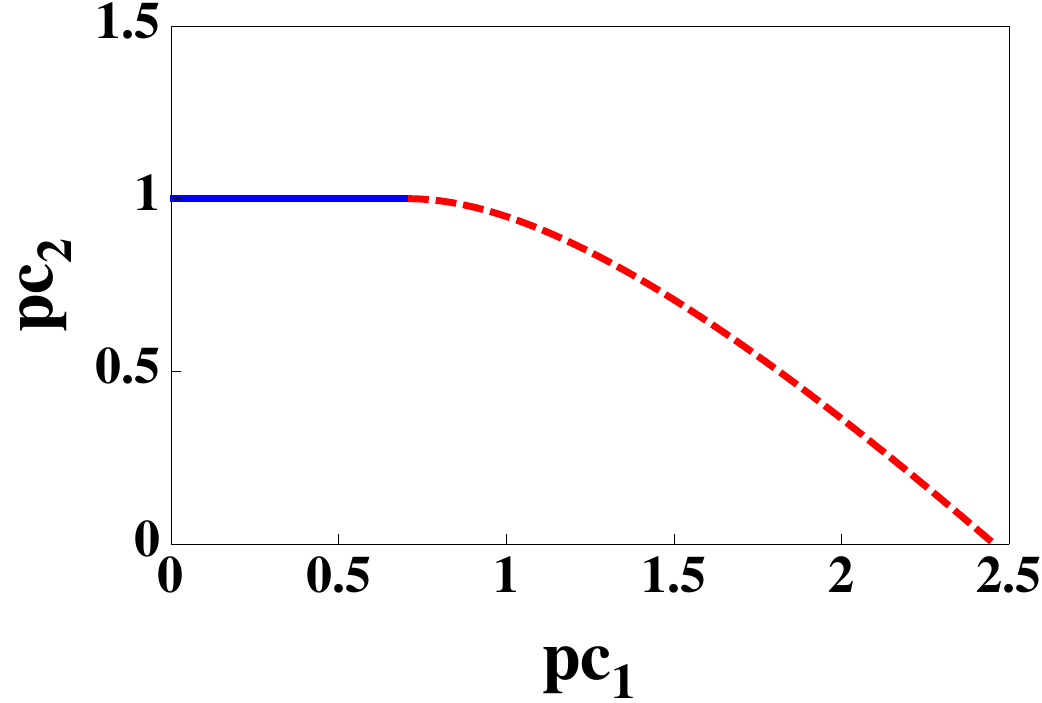}
	\caption{(Color online) The critical lines of discontinuous hybrid phase transition (red dashed line) and of continuous phase transition (blue solid line) describing the emergence of the DMCGC are shown 
	for the case of a multiplex networks with two layers and the Poisson multidegree distribution with  $\avg{k^{(1,0)}}=\avg{k^{(0,1)}}=c_1$ and $\avg{k^{(1,1)}}=c_2$.}
	\label{figDMGC2D}
\end{center}
\end{figure}

We consider now the case of a duplex $M=2$ in which the multidegree distributions are Poisson with   $\avg{k^{(1,1)}}=c_2$, and $\avg{k^{(0,1)}}=\avg{k^{(1,0)}}=c_1$.
Due to the properties of the Poisson distribution,  we have
$S=S_{\vec{m}}$, for every $\vec{m}\neq \vec{0}$, where $S$ satisfies the  equation
\begin{equation}
	S=p\left[1-2e^{-(c_1+c_2)S}+e^{-(2c_1+c_2)S}\right].
	\label{int_simple}
\end{equation}
By setting $x=S/p$, and $\hat{c}_1=c_1p,\hat{c_2}=c_2p$ we can study the solutions of the equivalent equation
\begin{equation}
	f(x)=x-\left[1-2e^{-(\hat{c}_1+\hat{c}_2)x}+e^{-(2\hat{c}_1+\hat{c}_2)x}\right]=0
\end{equation}
in the  $(\hat{c}_1,\hat{c}_2)$ parameter plane.
The critical line of discontinuous hybrid transition is found by solving the system of equations 
\bea
f(x_c)&=&0,\nonumber \\
\left.\frac{df(x)}{dx}\right|_{x=x_c}&=&0.
\eea
The critical line of second order phase transition is found by solving the equation 
\bea
\left. \frac{df(x)}{dx}\right|_{x=0}=0.
\eea
We notice that there is a non-trivial critical point for $c_2 p=1,c_2/c_1=\sqrt{2}$ for which 
\bea
\left. \frac{df(x)}{dx}\right|_{x=0}=\left. \frac{d^2f(x)}{dx^2}\right|_{x=0}0.
\eea
The full phase diagram of the model is displayed in Fig.~\ref{figDMGC2D}.
We note that for $c_2=0$  the transition is hybrid and discontinuous and reduces to the know transition in  duplex network with no link overlap, while for $c_1=0$ of complete overlap of the layers the transition is continuous and reduces to the percolation transition on a single Poisson network. 
\subsection{ Three  Poisson Layers with Overlap}

As a second example, we consider the ensemble of a three layer multiplex network $(M=3)$ with a Poisson multi-degree distribution and $\avg{k^{(1,0,0)}}=\avg{k^{(0,0,1)}}=\avg{k^{(0,1,0)}}=c_1$, $\avg{k^{(1,1,0)}}=\avg{k^{(1,0,1)}}=\avg{k^{(0,1,1)}}=c_2$ and $\avg{k^{(1,1,1)}}=c_3$.
As in the previous case, we have 
one order parameter $S=S_{\vec{m}}$  $\ \forall \vec{m}\neq {\vec{0}}$, and $S$  satisfies the equation
\begin{eqnarray}
	S &=& p \left[1-3e^{-(c_1+2c_2+c_3)S} \right.\nonumber\\
	&&+ \left. 3e^{-(2c_1+3c_2+c_3)S}-e^{-(3c_1+3c_2+c_3)S}\right].
	\label{3er-S-equation}
\end{eqnarray}
By setting  $x=S/p$, and $\hat{c}_1=c_1p,\hat{c}_2=c_2p,\hat{c}_3=c_3p$ we can define a function $g(x)$ as
\begin{eqnarray}
	g(x)&=&x-[1-3e^{-(\hat{c}_1+2\hat{c}_2+\hat{c}_3)x}+3e^{-(2\hat{c}_1+3\hat{c}_2+\hat{c}_3)x}\nonumber \\
	&& -e^{-(3\hat{c}_1+3\hat{c}_2+\hat{c}_3)x}]=0,
	\label{gx}
\end{eqnarray}
and we can recast the equation for $S$ as $g(x)=0$.

The phase diagram of this directed percolation problem is very rich. 
It includes non-trivial tricritical points. We refer the interested reader to the paper \cite{Cellai2013} that  investigate this case in detail.
Additionally we observe here that for $c_2=c_3=0$ we recover the prediction of the percolation transition in interdependent multiplex network with no link overlap, while for $c_1=c_2=0$ we recover the results of the percolation transition on a single Poisson network.

\section{Percolation with link overlap}
\label{sec:MCGC}

\subsection{General observations on the message passing algorithm for the MCGC}

Message passing algorithms are  powerful methods,
 allowing to solve exactly graphical models on locally tree-like networks, {\ie}  networks with a vanishing density of finite cycles. 
These methods are 
versatile, as they can be 
applied not only 
to ensembles of random networks, but  
also to single network realizations. 
For these reasons 
these algorithms are becoming increasingly popular 
in 
network science  
with applications ranging from percolation on single 
and multilayer networks 
to controllability \cite{Lenka,Kabashima,Son,BD1,BD2,Radicchi,Control}.
These algorithms proceed by iteration of dynamical rules which determine messages or beliefs that a node sent to a neighboring node.
In general, these beliefs indicate the probability that the neighbor node is in a given dynamical state and take real variables between zero and one.
Percolation on single networks, as well as  generalized percolation problems defined for multiplex networks, are inherently optimization problems in which one aims at characterizing the giant component, which is the largest connected component satisfying a set of conditions.
In this case the messages polarize and take only value $0,1$, as we have already seen in the  cases discussed so far. 

Moreover, in general, messages sent from node $i$ to a downstream node $j$  take into account 
not only the states of the upstream nodes $\ell \neq j$ but also the state of node $j$. 
We show that the message passing algorithm that determines the nodes in the MCGC of a multiplex network with overlap has the following properties: it is polarized ({\ie} the messages take values $0,1$) and assumes that the downstream node belongs to the MCGC.
These two properties of the message passing algorithm are 
in agreement with the general definition of the message passing algorithm, nevertheless, due to the second property, the resulting algorithm for detecting the MCGC loses the epidemic spreading interpretation when compared to the algorithm used to detect the DMCGC.

\subsection{ The message passing algorithm}

\begin{figure}[htb]
\begin{center}
	\includegraphics[width=0.8\columnwidth]{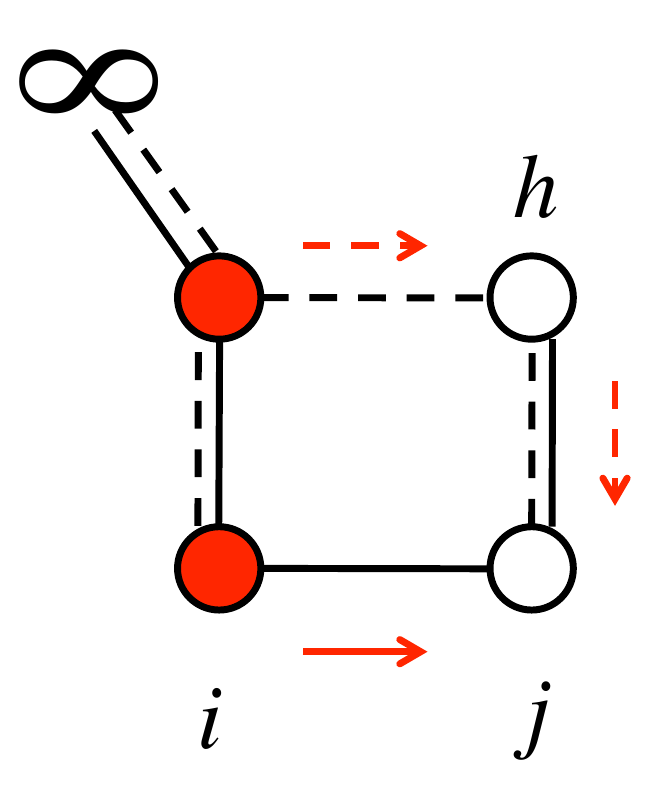}
	\caption{Non-directional character of the message passing approach to compute the MCGC.}
	\label{fig-MCGC-messages}
\end{center}
\end{figure}

Let $s_i=0,1$ indicate if a node is removed or not from the network and let   $\sigma_i=0,1$ be the indicator function that the node $i$ is in the mutually connected giant component. 
Let also $\vec{n}=(n_1,n_2,\ldots, n_M)$ be a fixed vector with $n_{\alpha}=0,1$.
The rationale of introducing vector $\vec{n}$ can be explained with the help of Fig.~\ref{fig-MCGC-messages} that 
highlights the difference between DMCGC and MCGC \cite{note}. 
In order to evaluate if node $j$ is in the MCGC, 
it is not sufficient to assume only  
the information upstream from node $i$. 
In fact, we must encode a system of messages that may reach node $j$ through different paths, on different layers.
In the network of Fig.~\ref{fig-MCGC-messages}, node $j$ belongs to the MCGC because it is reached by a positive message on layer 1 (solid) through node $i$ and by a positive message on layer 2 (dashed) through node $h$.
Therefore, our aim is to define the minimal set of layers that allows node $i$ to connect node $j$ to the MCGC through node $i$.
For a given multilink $\vec{m}^{ij}$, 
we will define below a vector $\vec{n}=\vec{n}^{i\to j}$ that encodes this selected set of layers.

Let us formally define  
our message passing algorithm that determines if a node is in the MCGC in a locally tree-like multiplex network.
The value of $\sigma_i$ is determined by the ``messages'' that the neighboring nodes send to node $i$. 
We denote the generic  ``message" as $\sigma_{i\to j}^{\vec{m}^{ij},\vec{n}}$. 
These messages are defined for every possible value of $\vec{n}$, and 
only for $\vec{n}=\vec{n}^{i\to j}$ we will have 
the message $\sigma_{i\to j}^{\vec{m}^{ij},\vec{n}^{i\to j}}=1$ indicating that node $i$ connects node $j$ to the MCGC exclusively through the links where $n_{\alpha}^{i\to j}=1$.

Let us now treat separately the cases $\vec{n}\neq \vec{0}$ and $\vec{n}=\vec{0}$.
For $\vec{n}\neq \vec{0}$  the value of the message $\sigma_{i\to j}^{\vec{m}^{ij},\vec{n}}$ is set to one, $\sigma_{i\to j}^{\vec{m}^{ij},\vec{n}}=1$, if and only if the following three conditions are satisfied: 

\begin{itemize}

	\item[(i)] 
	node $j$ is a neighbor of node $i$ with a multilink $\vec{m}^{ij}$ connecting them such that $\sum_{\alpha}m_{\alpha}^{ij}>0$;
	
	\item[(ii)] 
	assuming node $j$ belongs to  the mutually connected giant component, node $i$ is in the mutually connected component, {\ie} it is not initially damaged and it has at least one neighbor in any layer that belongs to the MCGC; 
	
	\item[(iii)] 
	node $i$ connects node $j$  to the  mutually connected component  \emph{exclusively} through  the  layers $\alpha$ with $n_{\alpha}=1$. 
	This implies that, if $m_{\alpha}^{ij}=0$, then this condition can only be satisfied if  $n_{\alpha}=0$, because otherwise we would need a link $m_{\alpha}^{ij}=1$ to allow for node $j$ to be connected to the MCGC on layer $\alpha$ through node $i$. 
	
\end{itemize}

If these three conditions are not met, and $\vec{n}\neq\vec{0}$,  we will have $\sigma_{i\to j}^{\vec{m}^{ij},\vec{n}}=0$.

An important consequence of this definition is the following.  
As per condition (iii), having a non-zero message  $\sigma_{i\to j}^{\vec{m}^{ij},\vec{n}}=1$ with $\vec{n}\neq \vec{0}$ requires that node $i$ connects node $j$  to the  (unique) MCGC \emph{exclusively} through  the  layers $\alpha$ with $n_{\alpha}=1$.
Then, there can be at most a single vector $\vec{n}\neq \vec{0}$ such that $\sigma_{i\to j}^{\vec{m}^{ij},\vec{n}}=1$. 
This vector, if it exists, will encode all the information about all the messages $\sigma_{i\to j}^{\vec{m}^{ij},\vec{n}}$ with $\vec{n}\neq \vec{0}$ and will indicate the set of layers that connect node $j$ to the MCGC through node $i$.
In order to treat at the same level the case in which node $i$ connects node $j$ to the MCGC at least in one layer and the case in which node $i$ does not connect node $j$ to the MCGC in any layer, it is  convenient to define the  messages $\sigma_{i\to j}^{\vec{m}^{ij},\vec{0}}$ as it follows:
\bea
\sigma_{i\to j}^{\vec{m}^{ij},\vec{0}}=\delta\left(0,\sum_{\vec{n}\neq \vec{0}}\sigma_{i\to j}^{\vec{m}^{ij},\vec{n}}\right),
\label{S0}
\eea
where $\delta(x,y)$ is the Kronecker function. 
Using this definition, we can define the unique vector $\vec{n}=\vec{n}^{i\to j}$ for which
\bea
\sigma_{i\to j}^{\vec{m}^{ij},\vec{n}^{i\to j}}=1,
\eea
as
\bea
n^{i\to j}=\mbox{argmax}_{\vec{n}}\sigma_{j\to i}^{\vec{m}^{ij},\vec{n}}.
\label{argn}
\eea
Therefore, the vector $\vec{n}^{i\to j}$ is uniquely defined for every linked (ordered) pair of nodes $(i,j)$ and its  components $\vec{n}^{i\to j}_{\alpha}$    indicate whether node $j$ is connected to the MCGC through node $i$ in layer $\alpha$ ($\vec{n}^{i\to j}_{\alpha}=1$) or not ($\vec{n}^{i\to j}_{\alpha}=0$).

Let us now derive the algorithm defining the messages.
We first consider  
condition (ii). Node $i$ is in the MCGC if and only if it is not initially damaged and it has at least one nearest neighbor in each layer $\alpha$ that  it is connected to  the MCGC.
Assuming that node $j$ is in the MCGC, this implies that node $i$ should have  in each layer $\alpha$ for which $m_{\alpha}^{ij}=0$, at least a neighbor different from {node $j$} that belongs to the MCGC.
Secondly, we consider condition (iii). According to this condition node $i$ connects node $j$ to the MCGC exclusively through layers $\alpha$ for which $n_{\alpha}=1$. Therefore   node $i$  should have  at least a neighbor different from node $j$ that belongs to the MCGC in each layer $\alpha$ for which $n_{\alpha}=1$, and should not have any neighbor different from node $j$ belonging to the  MCGC in the layers $\alpha$ for which $n_{\alpha}=0$.
Therefore the message $\sigma_{i\to j}^{\vec{m}^{ij},\vec{n}}$ with $\vec{n}\neq \vec{0}$ is equal to one, $\sigma_{i\to j}^{\vec{m}^{ij},\vec{n}}=1$, if and only if 
\begin{itemize}

\item[(a)] node $j$ is a neighbor of node $i$ with a multilink $\vec{m}^{ij}$ connecting them such that $\sum_{\alpha}m_{\alpha}^{ij}>0$;

\item[(b)] node $i$ is not initially damaged;

\item[(c)] for every layer $\alpha$  for which either $n_{\alpha}=1$ or  $m_{\alpha}^{ij}=0$ there is at least  
one node $\ell\neq j$, neighbor of node $i$ in layer $\alpha$, 
for which  $n_{\alpha}^{\ell\to i}=1$; 

\item[(d)] for every layer $\alpha$ for which  $m_{\alpha}^{ij}=1$ and $n_{\alpha}=0$  every node $\ell\neq j$, neighbor of node $i$ in layer $\alpha $, 
is sending a message with   $n_{\alpha}^{\ell\to i}=0$.

\end{itemize}
\begin{figure}[htb]
\begin{center}
	\includegraphics[width=1.0\columnwidth]{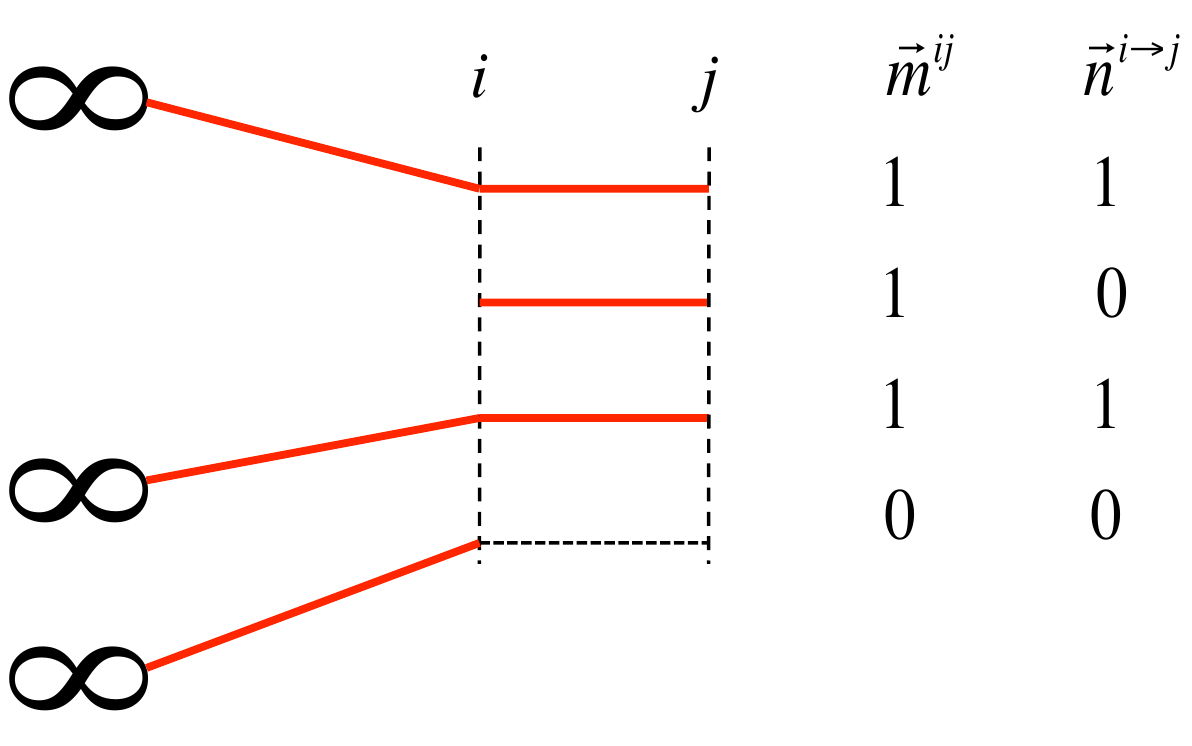}
	\caption{Schematical representation of the message passing algorithm to calculate the MCGC in the presence of link overlap. The $\infty$ symbols represent whether node $i$ is connected or not to the MCGC by a node different from $j$ on a given layer.}
	\label{fig-MCGC-algorithm}
\end{center}
\end{figure}
An example of this algorithm is schematically sketched in Fig.~\ref{fig-MCGC-algorithm}.

Therefore the message passing equation for the messages $ \sigma_{i\to j}^{\vec{m}^{ij},\vec{n}}$ reads 
 \bea
 \sigma_{i\to j}^{\vec{m}^{ij},\vec{n}}&=& s_i \prod_{\alpha=1}^M \left[1- \prod_{\ell \in N(i) \setminus j} \left(1-n_{\alpha}^{\ell \to  i}\right)\right]^{1-m_{\alpha}^{ij}(1-n_{\alpha})}\nonumber\\
 &&\times \left[ \prod_{\ell \in N(i) \setminus j} \left(1-n_{\alpha}^{\ell \to  i}\right)\right]^{m_{\alpha}^{ij}(1-n_{\alpha})}. 
 \label{Sp}
 \eea
 The value of  the indicator function $\sigma_i$  determining  if node $i$ belongs  ($\sigma_i=1$) or  not ($\sigma_i=0$) to the MCGC depends on the  messages $\sigma_{j\to i}^{\vec{m}^{ji},\vec{n}^{j\to i}}$. 
Specifically, node $i$ belongs to the MCGC if the following conditions are met:
\begin{itemize}

\item[(a)] 
node $i$ is not initially damaged, {\ie} $s_i=1$;

\item[(b)] 
node $i$ has at least one neighbor $j$ in each layer $\alpha$ that connects node $i$ to the MCGC, {\ie} for which $n_{\alpha}^{j\to i}=1$. 

\end{itemize}
Therefore we arrive at the message passing equations determining $\sigma_i$:  
\bea
\sigma_i=s_i \prod_{\alpha=1}^{M}\left[1-\prod_{j\in N(i)}\left(1-n_{\alpha}^{j \to i}\right)\right]
.
\label{S}
\eea

The message algorithm described above, consisting in iterating the Eqs. ($\ref{Sp}$),($\ref{argn}$), ($\ref{S0}$) and ($\ref{S}$) allows to predict which nodes of a  real locally tree-like multiplex network with link overlap are in the MCGC.
The message passing techniques are indeed guaranteed to converge to the correct solution only in locally tree-like networks, although empirically they may work surprisingly well on networks with some small cycles, as long as they have a vanishingly small clustering coefficient in the limit $N\to \infty$. Considering distributed algorithms that go beyond the tree-like approximation is a long standing challenge that has been tackled by the recent literature \cite{Radicchi_beyond}. Therefore, it can not be excluded that opportune variations of the algorithm that we have  introduced here for locally tree-like networks could be applied in the future to multiplex networks with finite clustering coefficient.
  
  In order to derive the equations for the average messages in  ensemble of multiplex networks chosen with probability given by Eq.~(\ref{PGM}) with  random initial damage of the nodes following  Eq.~(\ref{Ps}), let us now consider Eq.~(\ref{Sp}), by using the formula
\bea
	&&\prod_{\alpha=1}^M (1-x_{\alpha})^{p_{\alpha}}=\prod_{\alpha|p_{\alpha}>0}(1-z_{\alpha}) \nonumber \\
	&&= \sum_{\vec{r}|r_{\alpha}=0 \ \mbox{{\scriptsize if}}\  p_{\alpha}=0}(-1)^{\sum_{\alpha}r_{\alpha}} z_1^{r_1}\dots z_M^{r_M},
	\label{r}
\eea
valid as long as  $p_{\alpha}=0,1$.
Here $\vec{r}=(r_1,r_2,\ldots, r_{\alpha},\ldots ,r_M)$ with $r_{\alpha}=0,1$,
and where the sum $\sum_{\vec{r}}$ is over all possible vectors $\vec{r}$ with $r_{\alpha}=0$ in the layers $\alpha$ where  $p_{\alpha}=0$.
Since for each node $\ell$ {neighboring} node $i$ in layer $\alpha$ we should necessarily have ${m}^{\ell i}_{\alpha}=1$,  we can write
\bea
	\sigma_{i\to j}^{\vec{m}^{ij},\vec{n}} &=&s_i \sum_{\vec{r}|r_{\alpha}=0 \ \mbox{{\scriptsize if}}\  m_{\alpha}^{ij}(1-n_{\alpha})=1}(-1)^{\sum_{\alpha}r_{\alpha}} 
	\nonumber 
	\\
	&&\hspace{-5mm}\times\prod_{\alpha=1}^M\prod_{\ell \in N(i)\setminus j}\left(1-  n_{\alpha}^{\ell \to i}\right)^{[r_{\alpha}+m_{\alpha}^{ij}(1-n_{\alpha})]}
	.
	\eea	
	Since $n_{\alpha}^{\ell \to i}=0,1$ and $\sigma_{\ell \to i}^{\vec{m}^{\ell i},\vec{n}^{\ell \to i}}=1$,  we can write the above expression as
	\bea
	\sigma_{i\to j}^{\vec{m}^{ij},\vec{n}} &=&s_i \sum_{\vec{r}|r_{\alpha}=0 \ \mbox{{\scriptsize if}}\  m_{\alpha}^{ij}(1-n_{\alpha})=1}(-1)^{\sum_{\alpha}r_{\alpha}}
	\nonumber 
	\\
	&&
	\!\!\!\!\!\!\!\!\!\!\!\!\!\!\!\!\!\!\!\!\!\!\!\!\!\!\!\!\!\!\!\!
	\times \prod_{\alpha=1}^M \prod_{\ell \in N(i)\setminus j}\left(1- \sigma_{\ell\to i}^{\vec{m}^{\ell i},\vec{n^{\ell \to i}}} \right)^{n_{\alpha}^{\ell \to i}[r_{\alpha}+m_{\alpha}^{ij}(1-n_{\alpha})]}
	.
	\eea
	Finally this expression can be written as
	\bea
	\sigma_{i\to j}^{\vec{m}^{ij},\vec{n}} &=&s_i \sum_{\vec{r}|r_{\alpha}=0 \ \mbox{{\scriptsize if}}\  m_{\alpha}^{ij}(1-n_{\alpha})=1}(-1)^{\sum_{\alpha}r_{\alpha}} 
	\nonumber 
	\\
	&&
	\!\!\!\!\!\!\!\!\!\!\!\!\!\!\!\!\!\!\!\!\!\!\!\!\!\!\!\!\!\!\!\!
	\times\prod_{\ell \in N(i)\setminus j}\left(1- \sigma_{\ell\to i}^{\vec{m}^{\ell i},\vec{n^{\ell \to i}}} \right)^{\sum_{\alpha}n_{\alpha}^{\ell \to i}[r_{\alpha}+m_{\alpha}^{ij}(1-n_{\alpha})]}
	.
      		\label{sr}
		\eea
		
	Averaging over the ensemble of multiplex networks where each multiplex network is chosen with probability given by Eq.~(\ref{PGM}) and the random initial damage with probability given by Eq.~(\ref{Ps}), we get
\bea
\hspace{-0pt}S_{\vec{m},\vec{n}}\!\!&=&\!\!p\!\sum_{\{k^{\vec{m}}\}}\frac{k^{\vec{m}}}{\Avg{k^{\vec{m}}}}P(\{k^{\vec{m}}\})
\!\!\!\!\!\sum_{\vec{r}|r_{\alpha}=0 \ \mbox{{\scriptsize if}}\  m_{\alpha}(1-n_{\alpha})=1}
\!\!\!\!\!(-1)^{\sum_{\alpha=1}^M r_{\alpha}}
\nonumber 
\\
&&\hspace{-5mm}\times \prod_{\vec{m'}\neq \vec{m}}\left(1-\sum_{\vec{n'}|\sum_{\alpha} n_{\alpha}'[r_{\alpha}+m_{\alpha}(1-n_{\alpha})]>0}S_{\vec{m}',\vec{n'}}\right)^{k^{\vec{m'}}} \nonumber \\
&&\hspace{-5mm}\times \left(1-\sum_{\vec{n}'| \sum_{\alpha} n_{\alpha}'[r_{\alpha}+m_{\alpha}(1-n_{\alpha})]>0}S_{\vec{m}',\vec{n}'}\right)^{k^{\vec{m}}-1}
\eea
with $\vec{m}\neq \vec{0}$.	For networks with an uncorrelated multidegree distribution $P(\{k^{\vec{m}}\})$ given by Eq.~(\ref{UPGM}) we get 
\begin{eqnarray}
S_{\vec{m},\vec{n}}&=&p\sum_{\vec{r}|r_{\alpha}=0 \ \mbox{{\scriptsize if}}\  m_{\alpha}(1-n_{\alpha})=1}(-1)^{\sum_{\alpha=1}^M r_{\alpha}}
\nonumber 
\\
&&\hspace{-5mm}\times\prod_{\vec{m'}\neq \vec{m}}G^0_{\vec{m'}}\left(1-\sum_{\vec{n'}|\sum_{\alpha} n_{\alpha}'[r_{\alpha}+m_{\alpha}(1-n_{\alpha})]>0}S_{\vec{m}',\vec{n'}}\right) \nonumber \\
&&\hspace{-5mm}\times G^1_{\vec{m}}\left(1-\sum_{\vec{n}'| \sum_{\alpha} n_{\alpha}'[r_{\alpha}+m_{\alpha}(1-n_{\alpha})]>0}S_{\vec{m}',\vec{n}'}\right) \label{SnMCGC}
\eea
with $\vec{m}\neq \vec{0}$.\\
Similarly in order to derive the equation for $S=\avg{\sigma_i}$ let us now transform Eq. (\ref{S}) by using the formula
\begin{equation*}
	\prod_{\alpha=1}^M (1-z_{\alpha}) = \sum_{\vec{r}}(-1)^{\sum_{\alpha}r_{\alpha}} z_1^{r_1}\dots z_M^{r_M},
\end{equation*}
where $\vec{r}=(r_1,r_2,\ldots, r_{\alpha},\ldots r_M)$ with $r_{\alpha}=0,1$,
and where the sum $\sum_{\vec{r}}$ is over all possible vectors $\vec{r}$. We expand the multiplications and write
\bea
	\sigma_{i} &=& s_i\sum_{\vec{r}}(-1)^{\sum_{\alpha}r_{\alpha}}\prod_{\alpha=1}^M\prod_{j\in N(i)}\left(1-n_{\alpha}^{j\to i}\right)^{r_{\alpha}}
	.
	\eea
Since $n_{\alpha}^{j\to i}=0,1$ and $\sigma_{j \to i}^{\vec{m}^{ji},\vec{n}^{j\to i}}=1$,  we can write the above expression as
	\bea
	&&
	\!\!\!\!\!\!\!\!\!
	\sigma_i=s_i\sum_{\vec{r}}(-1)^{\sum_{\alpha}r_{\alpha}}\prod_{\alpha=1}^M\prod_{j\in N(i)}\left(1-\sigma_{j \to i}^{\vec{m}^{ji},\vec{n}^{j\to i}}\right)^{r_{\alpha}n_{\alpha}^{j\to i}}
	\nonumber 
	\\
	&&
	\!\!\!\!\!\!\!\!\!\!
	=s_i\sum_{\vec{r}}(-1)^{\sum_{\alpha}r_{\alpha}}\prod_{j\in N(i)}\left(1-\sigma_{j \to i}^{\vec{m}^{ji},\vec{n}^{j\to i}}\right)^{\sum_{\alpha}r_{\alpha}n_{\alpha}^{j\to i}}
	.
	\label{sigma2}
\eea

We average this expression over the ensemble of network by choosing a multiplex network with probability given by Eq.~(\ref{PGM}) and averaging over the random realization of the initial damage according to the probability distribution in Eq.~({\ref{Ps}}). In this way we get the expression for the average number of nodes $S$ in the MCGC:  
\begin{eqnarray}
S&=&p\sum_{\{k^{\vec{m}}\}}P(\{k^{\vec{m}}\})\sum_{\vec{r}}(-1)^{\sum_{\alpha=1}^M r_{\alpha}}
\nonumber 
\\
&&\times\prod_{\vec{m}}\left(1-\sum_{\vec{n}|\sum_{\alpha} n_{\alpha}r_{\alpha}>0}S_{\vec{m},\vec{n}}\right)^{k^{\vec{m}}}
,
\eea
where  $S_{\vec{m},\vec{n}}=\Avg{\sigma_{j \to i}^{\vec{m}^{ji},\vec{n}}}$ is the probability  that a node  is connected through a multilink $\vec{m}$  to the mutually connected giant component through the layers indicated by the vector $\vec{n}$. 
In particular, in an ensemble in which the multi-degree distribution factorizes, {\ie} the multidegree distribution $P(\{k^{\vec{m}}\})$  follows Eq.~(\ref{UPGM}), 
we have 
\bea
S&=&p\sum_{\{k^{\vec{m}}\}}\sum_{\vec{r}}(-1)^{\sum_{\alpha=1}^M r_{\alpha}}
\nonumber 
\\
&&\times\prod_{\vec{m}}G^0_{\vec{m}}\left(1-\sum_{\vec{n}|\sum_{\alpha} n_{\alpha}r_{\alpha}>0}S_{\vec{m},\vec{n}}\right).
\label{SMCGC}
\eea

Notice 
that the Eq. ($\ref{SnMCGC}$) and Eq. $(\ref{SMCGC})$ in the case of networks without link overlap, {\ie} where all multilinks $\vec{m}\neq{0}$ have $\sum_{\alpha}m_{\alpha}=1$, reduce to the equations found in Sec. \ref{sec:percolation-no-overlap}. In the following we consider two cases of multiplex networks with non-trivial link overlap formed respectively by two and three layers. 

\begin{figure}
\begin{center}
	\includegraphics[width=1.0\columnwidth]{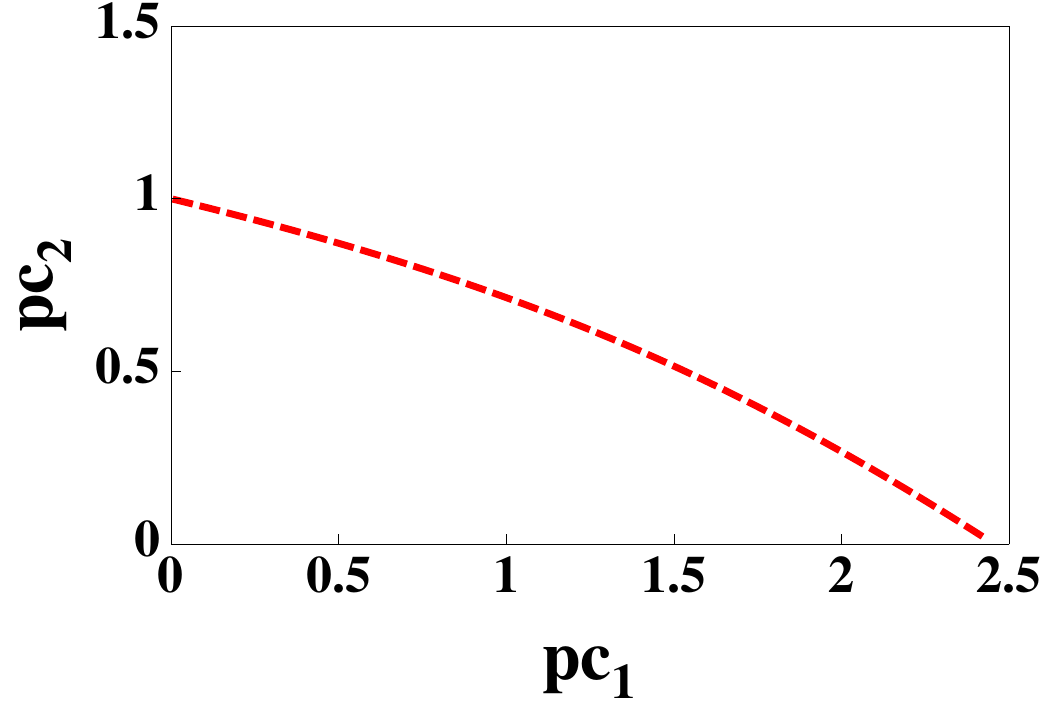}
	\caption{(Color online) The critical line of a discontinuous hybrid phase transition is shown 
for 
a multiplex networks with two layers and Poisson multidegree distribution with  $\avg{k^{(1,0)}}=\avg{k^{(0,1)}}=c_1$ with $\avg{k^{(1,1)}}=c_2$.}
	\label{figMCGC2D}
\end{center}
\end{figure}

\subsection{Two Layers with Overlap }
\label{m2}

As a first example, we consider here a multiplex networks formed by  two  layers with Poisson multidegree distribution and  $\avg{k^{(1,0)}}=\avg{k^{(0,1)}}=c_1$ with $\avg{k^{(1,1)}}=c_2$.  Here we show that the message passing approach proposed in this paper reproduces the same equations found in \cite{Baxter2016} and in total agreement with simulations results \cite{Baxter2016,supernodes}.
In this case the dynamical  variables that we have to consider are 
\bea
S=S_{\vec{m},\vec{m}}=x/p,
\eea
with $\vec{m}\neq \vec{0}$
\bea
S_{\vec{1},{(1,0)}}=S_{\vec{1},(0,1)}=x_{2,1}/p.
\eea
The  Eqs.~(\ref{SMCGC}), (\ref{SnMCGC}) for $x$ and $x_{2,1}$ read  
\bea
\left(\begin{array}{c}F_1({\bf x})
\nonumber 
\\ 
F_2({\bf x})\end{array}\right)={\bf F}({\bf x})={\bf 0}, 
\eea
where the functions $F_1({\bf x})$ and $F_2{(\bf{x})}$ are given by 
\bea
F_1(\bf{x})\!\!&=&\!\!x{-}\!\left(1{-}2e^{-\hat{c}_1x-\hat{c}_2(x+x_{2,1})}{+}e^{-2\hat{c}_1x-\hat{c}_2(x+2x_{2,1})}\right),\nonumber 
\\
F_2(\bf{x})\!\!&=\!\!&u{-}\!\left(e^{-\hat{c}_1x-\hat{c}_2(x+x_{2,1})}{-}e^{-2\hat{c}_1x-\hat{c}_2(x+2x_{2,1}) }\right),
\eea
the vector ${\bf x}$ is given by 
 \bea
{\bf x}=\left(\begin{array}{c}x\nonumber \\ x_{2,1}\end{array}\right),
\eea
and  $\hat{c_1}=c_1p$ and $\hat{c}_2=c_2p$. As expected, these equations are equivalent to the ones derived with the tree-like approximation in Ref.~\cite{Baxter2016}.

The points of  discontinuous hybrid phase transition can be found by imposing the set of equations 
\bea
{\bf F}({\bf x^{\star}})={\bf 0}\nonumber \\
\left.\det{\bf J}\right|_{{\bf x=x^{\star}}}=0,
\eea
where ${\bf J}$ is the Jacobian matrix of ${\bf F}(\bf{x})$.
The critical point of   continuous phase transition can be found by imposing 
\bea
\left.\det{\bf J}\right|_{{\bf x=0}}=0.
\eea
This equation can be expressed explicitly as
\bea
1 - 2 \hat{c}_2 + \hat{c}_2^2=0
\eea
and  has a unique real solution  for $\hat{c}_2=1$.
Analyzing these equations provides the full phase diagram of the model displayed in 
Fig.~\ref{figMCGC2D}.
The MCGC component emerges as a continuous phase tradition only when all links overlap, {\ie} $c_2p=1, c_1=0$ when we recover the case of percolation in a single Poisson network.
Finally we observe that for $c_2=0$ we recover the known results of percolation transition in interdependent duplex Poisson network with no link overlap.

\subsection{Three Layers with Overlap }
\label{m3}

As a second example, we consider a multiplex network formed by  three  layers with a Poisson multidegree distribution and  $\avg{k^{(1,0,0)}}=\avg{k^{(0,1,0}}=\avg{k^{(0,0,1)}}=c_1$ with $\avg{k^{(1,1,0)}}=\avg{k^{(1,0,1)}}=\avg{k^{(0,1,1)}}=c_2$ and $\avg{k^{(1,1,1)}}=c_3$. This case provides an example of a three layers multiplex network with overlap. The MCGC on this class of networks has never been solved with previous methods and therefore it demonstrated that the present theory allows to go beyond the previously available theoretical methods and techniques. 
In this case the  dynamical variables determining the percolation transition are 
\bea
S=S_{\vec{m},\vec{m}}=x/p,
\eea
with $\vec{m}\neq \vec{0},\vec{1}{=}(1,1,1)$, 
\bea
&&
\!\!\!\!\!\!\!
S_{\vec{1},{(1,1,0)}}=S_{\vec{1},(0,1,1)}=S_{\vec{1},(1,0,1)}=x_{3,2}/p,
\nonumber 
\\
&&
\!\!\!\!\!\!\!
S_{\vec{1},{(1,0,0)}}=S_{\vec{1},(0,1,0)}=S_{\vec{1},(0,0,1)}=x_{3,1}/p.
\nonumber 
\\
&&
\!\!\!\!\!\!\!
S_{{(1,1,0)},{(1,0,0)}}=S_{(1,1,0),(0,1,0)}=S_{(0,1,1),(0,1,0)}
\nonumber
\\
&&
\!\!\!\!\!\!\!
=S_{(0,1,1),(0,0,1)}=S_{(1,0,1),(1,0,0)}=S_{(1,0,1),(0,0,1)}{=}x_{2,1}/p.
\nonumber 
\eea

\begin{figure}
\begin{center}
	\includegraphics[width=1.0\columnwidth]{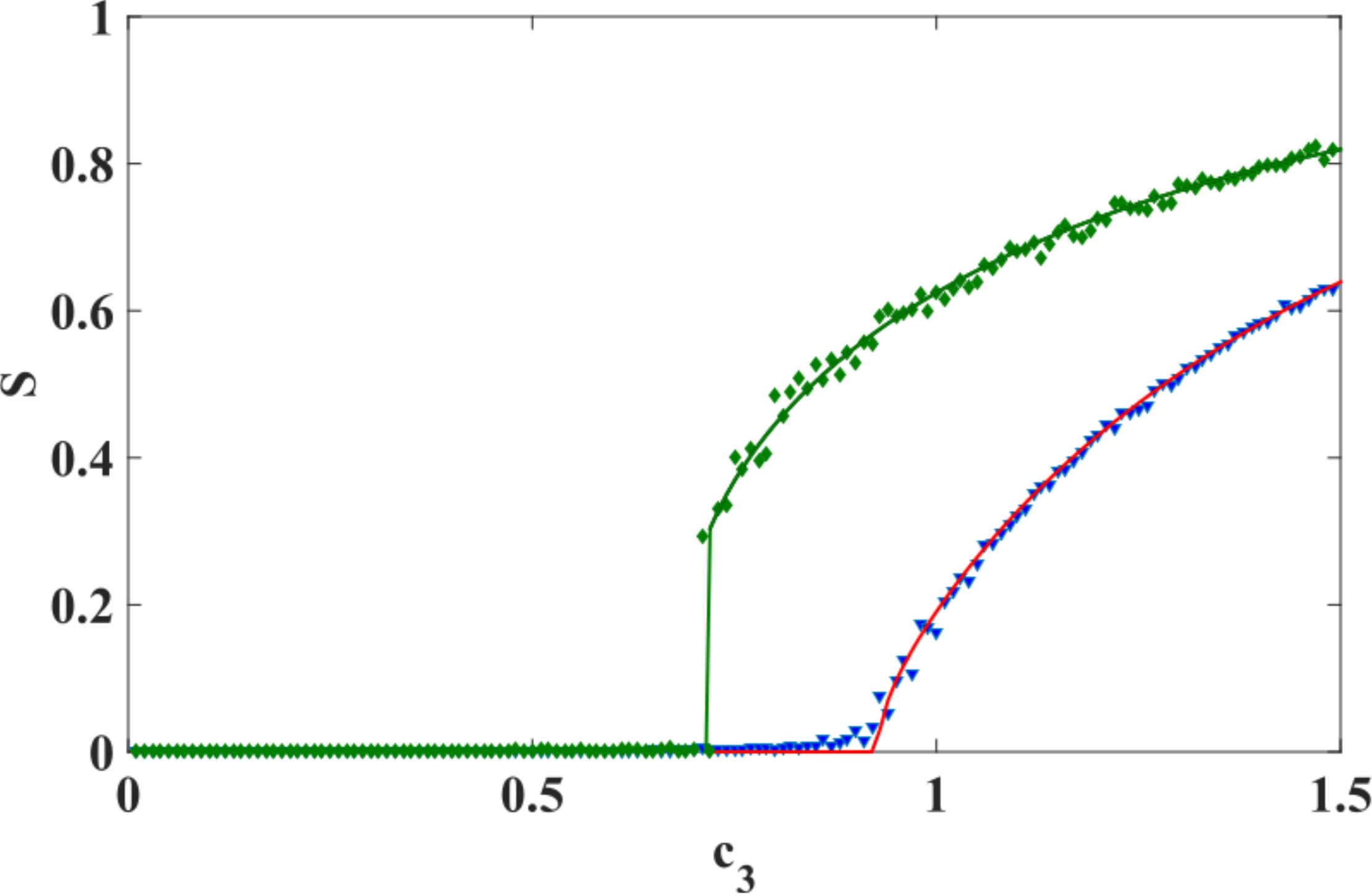}
	\caption{(Color online) Simulations of the MCGC for $p=1$ are shown 
	as a function of $c_3$ for $c_1=1.0,c_2=0.15$ (green diamonds) and for $c_1=0.4,c_2=0.0$ (blue triangles). The results were obtained from simulation of $3$-layer multiplex networks with $N=10^4$ nodes.  The data have been obtained for a single realization in the case of $c_1=1.0,c_2=0.15$, and they have been averaged over $10$ realizations for $c_1=0.4,c_2=0.0$.The simulations results perfectly match the theoretical expectations (solid lines). Notice that in both case we have a discontinuous jump although the jump is too small to be appreciated for $c_1=0.4,c_2=0.0$.}
	\label{figsimulations}
\end{center}
\end{figure}

\begin{figure}
\begin{center}
	\includegraphics[width=1.0\columnwidth]{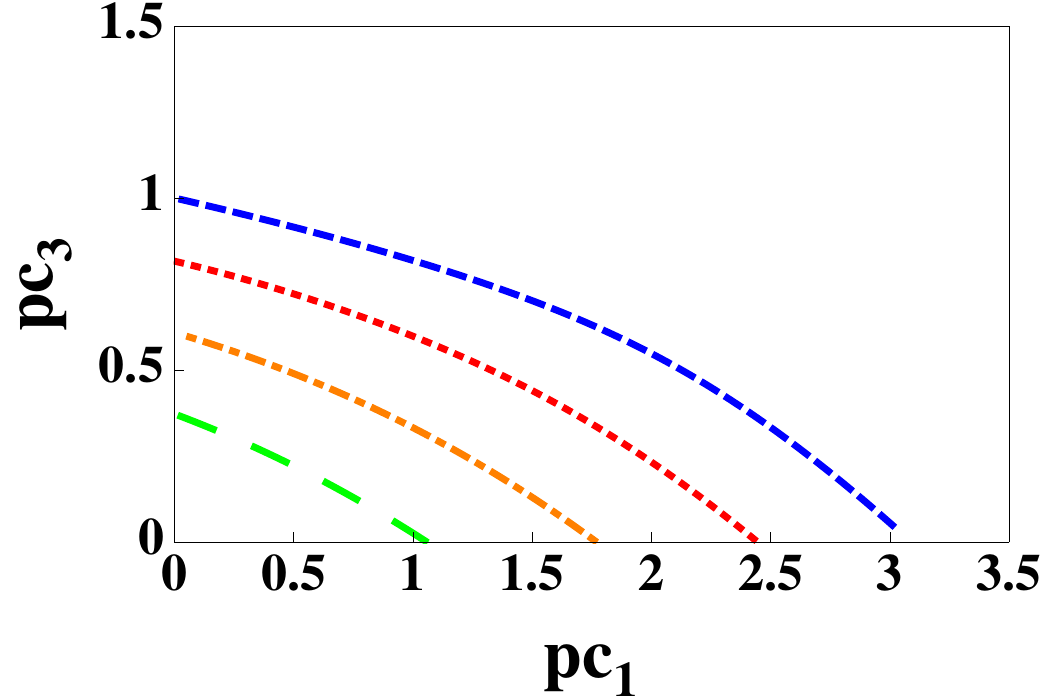}
	\caption{(Color online) The lines of critical points for the discontinuous hybrid transition describing the emergence of the MCGC are 
	shown for the case of  a multiplex networks formed layers with a Poisson multidegree distribution with   $\avg{k^{(1,0,0)}}=\avg{k^{(0,1,0}}=\avg{k^{(0,0,1)}}=c_1$ with $\avg{k^{(1,1,0)}}=\avg{k^{(1,0,1)}}=\avg{k^{(0,1,1)}}=c_2$ and $\avg{k^{(1,1,1)}}=c_3$.
 The  lines of the figure refer to critical lines for constant values of $pc_2$ given respectively by $pc_2=0.0$ (blue dashed line), $pc_2=0.25$ (red dotted line) $pc_2=0.5$ (orange dot-dashed line) and $pc_2=0.75$ (green long-dashed line).}
	\label{figMCGC3D_c2}
\end{center}
\end{figure}

By setting $\hat{c}_1=c_1p, \hat{c}_2=c_2p$ and $\hat{c}_3=c_3p$, Eqs.~(\ref{SMCGC}), (\ref{SnMCGC}) for $x, x_{3,2},x_{3,1},x_{2,1}$ read
\bea
\left(\begin{array}{c}G_1({\bf x})\nonumber \\ G_2(\bf{x})\\ G_3({\bf x})\\G_4({\bf x})\end{array}\right)={\bf G}({\bf x})={\bf 0}, 
\eea
where the functions $G_{\mu}({\bf x})$ with $\mu=1,2,3,4$ are given by 

\bea
G_1({\bf x})&=&x-\left[1-3e^{-\hat{c}_1x-\hat{c}_2(2x+2x_{2,1})-\hat{c}_3(x+2x_{3,2}+x_{3,1})}\right.\nonumber 
\\
&&
\left.+3e^{-2\hat{c}_1x-\hat{c}_2(3x+4x_{2,1})-\hat{c}_3(x+3x_{3,2}+2x_{3,1})}\right.
\nonumber 
\\
&&
\left.-e^{-3\hat{c}_1x-3\hat{c}_2(x+2x_{2,1})-\hat{c}_3(x+3x_{3,2}+3x_{3,1})}\right]
,
\nonumber 
\\
G_2({\bf x})&=&x_{3,2}-e^{-\hat{c}_1x-\hat{c}_2(2x+2x_{2,1})-\hat{c}_3(x+2x_{3,2}+x_{3,1})}
\nonumber 
\\
&&
\left[1-2e^{-\hat{c}_1x-\hat{c}_2(x+2x_{2,1})-\hat{c}_3(x_{3,2}+x_{3,1})}\right.
\nonumber 
\\
&&
\left.+e^{-2\hat{c}_1x-\hat{c}_2(x+4 x_{2,1})-\hat{c}_3(x_{3,2}+2x_{3,1})}\right],
\nonumber 
\\
G_3({\bf x})&=&x_{3,1}-e^{-2\hat{c}_1x-\hat{c}_2(3x+4x_{2,1})-\hat{c}_3(x+3x_{3,2}+2x_{3,1})}
\nonumber 
\\
&&\times \left[1-e^{-\hat{c}_1x-2\hat{c}_2x_{2,1}-\hat{c}_3x_{3,1}}\right],
\nonumber 
\\
G_4({\bf x})&=&x_{2,1}-x_{3,2},
\eea
and
\bea
{\bf x}=\left(\begin{array}{c}x\nonumber \\ x_{3,2}\\x_{3,1}\\x_{2,1}\end{array}\right)
.
\eea

The points of discontinuous hybrid phase transition can be found 
from the set of equations 
\bea
{\bf G}({\bf x^{\star}})&=&{\bf 0}
,
\nonumber 
\\
\left.\det{\bf J}\right|_{{\bf x=x^{\star}}}&=&0,
\eea
where ${\bf J}$ is the Jacobian of ${\bf G}(\bf{x})$.
The point of continuous phase transition can be found 
from the condition: 
\bea
\left.\det{\bf J}\right|_{{\bf x=0}}=0.
\eea
This equation,  
\bea
&&1 - 3 \hat{c}_3 + 3 \hat{c}_3^2 - \hat{c}_3^3=0,
\eea
has a unique real  solution  for $ \hat{c}_3=1$.
Analyzing the phase diagram one can see that this continuous phase transition occurs only for ${c}_1={c}_2=0, c_3p=1$ recovering the result of percolation on single Poisson network.
Additionally, for  $c_2=c_3=0$ we recover the known results in absence of link overlap.
In Fig.~\ref{figMCGC3D_c2}  and in Fig.~\ref{figMCGC3D_c1} we report sections of the phase diagram at constant values of $\hat{c}_2$ and at constant values of $\hat{c}_1$, respectively.

We have checked these equations  against simulation results showing that the analytical results perfectly match the simulations as it is shown in Fig.~\ref{figsimulations}.
\begin{figure}
\begin{center}
	\includegraphics[width=1.0\columnwidth]{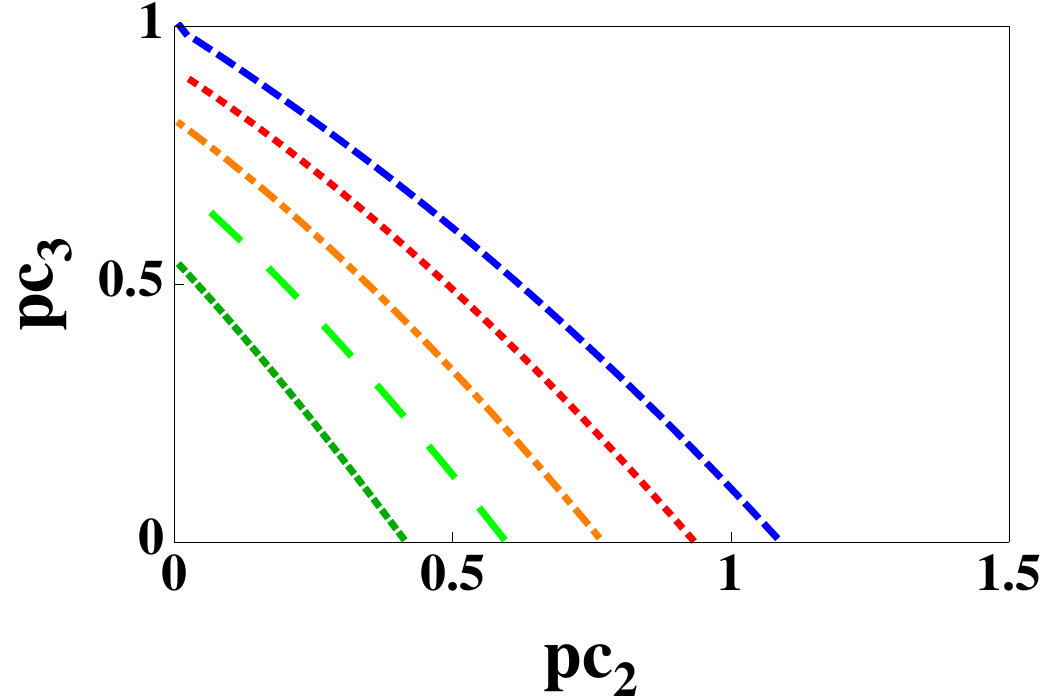}
	\caption{(Color online) The lines of critical points for the discontinuous hybrid transition describing the emergence of the MCGC are 
	shown for the case of  a multiplex network formed by three layers with  a Poisson multi degree distribution with   $\avg{k^{(1,0,0)}}=\avg{k^{(0,1,0}}=\avg{k^{(0,0,1)}}=c_1$ with $\avg{k^{(1,1,0)}}=\avg{k^{(1,0,1)}}=\avg{k^{(0,1,1)}}=c_2$ and $\avg{k^{(1,1,1)}}=c_3$.
 The  lines of the figure refer to critical lines for constant values of $pc_1$ given respectively by $pc_1=0.0$ (blue, dashed line), $pc_1=0.5$ (red dotted line) $pc_1=1.0$ (orange dot-dashed line), $pc_1=1.5$ (green long-dashed line) and $pc_1=2.0$ (dark green tiny-dashed line).}
	\label{figMCGC3D_c1}
\end{center}
\end{figure}

\section{Conclusions}
\label{sec:conclusions}

In this paper we have proposed a 
general unified message passing theory to calculate analytically mutual percolation on locally tree-like multiplex networks. While recent message passing methods had so far mostly dealt with multiplex networks without link overlap, here we have shown that this approach can be generalized to the latter, more difficult, case.
Our results show explicitly that one can describe the 
mutual component without resorting to super-nodes \cite{supernodes,Goh_comment},  which were used for investigating two-layer multiplexes with overlap.
Additionally, our approach allows the immediate treatment of the percolation transition in multiplex networks with an arbitrary number of layers $M$, extending greatly the variety of multiplex networks that can be studied.

Here we have distinguished between   two different percolation problems---directed and classical mutual percolation. 
These  percolation problems both reduce to the original mutual percolation scenario
if there is no link overlap. 
Our formalism shows that percolation and directed percolation in multiplex network with link overlap present different phase diagrams.
While the directed percolation transition describing the emergence of the directed mutually connect giant component is modeled by a feed-forward message passing algorithm that can mimic an epidemic spreading, as it was recently investigated \cite{Cellai2013}, the percolation transition is solved by a new message passing algorithm that does not have this feed-forward character.
We explored the transitions and the giant components in multiplex networks with two and three partially overlapping layers, for which we derived explicit equations. 
In a similar way, appropriate message passing algorithms can be used to determine the percolation transitions for any finite number of layers. 
We suggest that this version of the message passing approach can be successfully applied to even more complex multiplex networks.   
 
 \acknowledgments
 We acknowledge useful discussions with G.~J.~Baxter, R.~A.~da Costa, J.~P.~Gleeson, and J.~F.~F.~Mendes.  This work was partially supported by the FET proactive IP project MULTIPLEX 317532, Science Foundation Ireland, grant 14/IF/2461; the FET-Proactive project PLEXMATH (FP7-ICT-2011-8; grant 317614).


\begin{thebibliography}{99}

\bibitem{PhysReports}
S. Boccaletti, G. Bianconi, R. Criado, C. I. del Genio, J. G\'omez-Garde\~nes, M. Romance, I. Sendi\~na-Nadal, Z. Wang, and M. Zanin, Phys. Rep. {\bf 544}, 1 (2014). 
 
\bibitem{Kivela}
M. Kivel\"a, A.  Arenas, M. Barthelemy, J. P. Gleeson, Y. Moreno, and M. A. Porter,  
J. Complex Netw. {\bf 2}, 203 (2014). 

\bibitem{Goh_review}
K.-M.~Lee, B.~Min, and K.-I.~Goh,
 Eur. Phys. Jour. B {\bf 88}, 1 (2015).

\bibitem{Thurner}
M. Szell, R. Lambiotte, and S. Thurner, PNAS, {\bf 107}, 13636 (2010).

\bibitem{Mucha}
 P. J. Mucha, T. Richardson, K. Macon, M. A. Porter, and J.-P. Onnela,
Science, {\bf 328}, 876 (2010).

\bibitem{Bullmore2009} 
E. Bullmore and O. Sporns,  
{Nat. Rev. Neurosci.} \textbf{10}, 186 (2009).

\bibitem{Boccaletti}
A. Cardillo, J. G\'omez-Garde\~nes, M. Zanin, M. Romance, D. Papo, F. del Pozo, and S. Boccaletti, Sci. Rep. {\bf 3}, 1344 (2013).

\bibitem{Makse}
S. D. S. Reis, Y. Hu, A. Babino, J. S. Andrade Jr., S. Canals, M. Sigman, and H. A. Makse, 
Nature Phys. {\bf 10}, 762 (2014). 

\bibitem{Weighted}
G. Menichetti, D. Remondini, P. Panzarasa, R. J. Mondrag\'on,  and G. Bianconi, 
PloS one {\bf 9}, e97857 (2014).

\bibitem{Vito}
V. Nicosia and V. Latora, 
Phys. Rev. E {\bf 92}032805 (2015).

\bibitem{Havlin1}
S. V. Buldyrev, R. Parshani, G. Paul, H. E. Stanley, and S. Havlin, 
Nature {\bf  464}, 1025 (2010). 

\bibitem{Dorogovtsev}
G.~J. Baxter, S.~N. Dorogovtsev, A.~V. Goltsev, and J.~F.~F. Mendes, 
Phys. Rev. Lett. {\bf 109}, 248701 (2012). 

\bibitem{Son}
S.-W. Son, G. Bizhani, C. Christensen, P. Grassberger, and M. Paczuski, 
EPL {\bf 97}, 16006 (2012).

\bibitem{crit} 
S. N. Dorogovtsev, A. Goltsev, and J. F. F. Mendes, 
Rev. Mod. Phys. {\bf 80}, 1275 (2008).

\bibitem{Lenka}
B. Karrer, M. E. J. Newman, and L. Zdeborov\'a,
 Phys. Rev. Lett. {\bf 113},  208702 (2014).
 
 \bibitem{Mezard}
M. Mezard and A. Montanari, 
{\it Information, Physics and Computation} 
(Oxford University Press, Oxford, 2009).

\bibitem{Weigt}
A.~K. Hartmann and M.~Weigt, 
{\it Phase Transitions in Combinatorial Optimization Problems},
(WILEY-VCH, Weinheim, 2005). 

 \bibitem{Marianselfsimilar}
M. A. Serrano, D. Krioukov, and M. Bogu\~n\'a,
Phys. Rev. Lett. {\bf 106}, 048701 (2011).

 \bibitem{Radicchi_beyond}
F. Radicchi  and C. Castellano,
Phys. Rev. E {\bf 93},  030302 (2016).


\bibitem{Antagonist}
K. Zhao and G. Bianconi,  
J. Stat. Mech. P05005 (2013). 

\bibitem{Kun_q}
K. Zhao and G. Bianconi,
J. Stat. Phys. {\bf 152},  1069 (2013). 

\bibitem{Weak}
G. J. Baxter, S. N. Dorogovtsev, J. F. F. Mendes,  and D. Cellai,
Phys. Rev. E {\bf 89}, 042801 (2014). 

\bibitem{baxter2016unified}
G. J. Baxter, D. Cellai, S. N. Dorogovtsev, A. Goltsev, and J. F. F. Mendes, in Interconnected Networks, edited by A. Garas (Springer International Publishing), Understanding Complex Systems, 101 (2016).

\bibitem{Kcore}
N. Azimi-Tafreshi,  J. G\'omez-Garde\~nes,  and S. N. Dorogovtsev,  
Phys. Rev. E {\bf 90}, 032816 (2014).

\bibitem{Doro_directed}
N. Azimi-Tafreshi, S. N. Dorogovtsev, and J. F. F. Mendes,
Phys. Rev. E {\bf 90}, 052809 (2014).

\bibitem{Guha2016}
S. Guha, D. Towsley, P. Nain, \c{C}. \c{C}apar, A. Swami, and P. Basu
Phys. Rev. E {\bf 93}, 062310 (2016).

\bibitem{Bond}
A. Hackett, D. Cellai, S. G\'omez, A. Arenas, and J. P. Gleeson,
Physical Review X 6, 021002 (2016).


\bibitem{Kabashima}
S. Watanabe and Y. Kabashima, 
Phys. Rev. E. {\bf 89}, 012808 (2014).

\bibitem{Goh}
B. Min, S.~ D. Yi, K.-M. Lee, and K.-I. Goh, 
Phys. Rev. E {\bf 89}, 042811 (2014).

\bibitem{HavlinEPL}
R. Parshani, C. Rozenblat, D. Ietri, C. Ducruet, and S. Havlin, 
EPL \textbf{92}, 68002 (2010).

\bibitem{Havlin2}
R. Parshani, S. V. Buldyrev, and S. Havlin, 
Phys. Rev. Lett. {\bf 105}, 048701 (2010).

\bibitem{Stanleyint}
G. Dong, L. Tian, R. Du, J. Xiao, D. Zhou, and H.~E. Stanley, 
EPL {\bf 102}, 68004 (2013).

\bibitem{Cellai2016}
D. Cellai,  and G. Bianconi,
Phys. Rev. E {\bf 93}, 032302 (2016).

\bibitem{Gao1}
J. Gao, S. V. Buldyrev, H. E. Stanley, and S. Havlin, 
Nature Phys. {\bf 8}, 40 (2012).


\bibitem{BD1}
G. Bianconi, S.~N. Dorogovtsev, and J.~F.~F. Mendes, 
Physical Review E {\bf 91}, 012804 (2015).

\bibitem{BD2}
G. Bianconi and S.~N. Dorogovtsev,  
Phys. Rev. E {\bf 89}, 062814 (2014).



\bibitem{PRE}
G. Bianconi, 
Phys. Rev. E {\bf 87}, 062806 (2013).


\bibitem{note0}
The overlap of edges corresponds to the presence of correlations among 
different layers.

\bibitem{supernodes}
Y. Hu, D. Zhou, R. Zhang, Z. Han, C. Rozenblat,  and
S. Havlin, 
Phys. Rev. E {\bf 88}, 052805 (2013).

\bibitem{Goh_comment}
B. Min, S. Lee, K.-M. Lee, and K-I. Goh,
Chaos, Solitons \& Fractals {\bf 72} 49 (2015).

\bibitem{Baxter2016}
G. J. Baxter, G. Bianconi, R. A. da Costa, S. N. Dorogovtsev, and J. F. F.  Mendes,
Phys. Rev. E {\bf  94}, 012303 (2016).



\bibitem{Cellai2013}
D. Cellai, E. L\'opez, J. Zhou, J. P. Gleeson, and G. Bianconi,
Phys. Rev. E {\bf 88},  052811 (2013). 

\bibitem{Radicchi}
F. Radicchi, 
Nature Physics {\bf 11}, 597 (2015).



\bibitem{bootstrap}
G. J. Baxter, S. N. Dorogovtsev, A. V. Goltsev and J. F. F. Mendes,
Phys. Rev. E {\bf 82}, 011103 (2010). 

\bibitem{Azimi}
N. Azimi-Tafreshi, 
arXiv preprint arXiv:1511.03235 (2015).

\bibitem{Colizza}
E. Valdano,  L. Ferreri, C. Poletto, and V. Colizza,
Phys. Rev. X {\bf 5},  021005 (2015).

\bibitem{Control}
Y.Y. Liu,  J.-J. Slotine, and A.-L. Barab\'asi,
Nature {\bf 473}, 167 (2011).

\bibitem{Fakhteh}
W. Cai, L. Chen, F. Ghanbarnejad, and P. Grassberger,
Nature Physics {\bf 11}, 936 (2015).

\bibitem{note}
{This example 
is only for illustration purposes, as 
this 
is obviously not a locally tree-like network.}



\end{thebibliography}
\end{document}